\documentclass[11pt]{article}
\usepackage{amsmath}
\usepackage{amsfonts}
\usepackage{amssymb}
\usepackage[usenames,dvipsnames]{color}
\usepackage{epsf}
\usepackage{wick}

\newcommand{\be}{\begin{equation}}
\newcommand{\ee}{\end{equation}}
\newcommand{\ba}{\begin{eqnarray}}
\newcommand{\ea}{\end{eqnarray}}
\newcommand{\nn}{\nonumber}
\newcommand{\barr}{\begin{array}}
\newcommand{\earr}{\end{array}}

\newcommand\lsim{\mathrel{\rlap{\lower4pt\hbox{\hskip1pt$\sim$}}
        \raise1pt\hbox{$<$}}}
\newcommand\gsim{\mathrel{\rlap{\lower4pt\hbox{\hskip1pt$\sim$}}
        \raise1pt\hbox{$>$}}}

\def\x{{\bf x}}
\def\k{{\bf k}}
\def\q{{\bf q}}
\def\r{{\bf r}}
\def\fnl{f_{NL}}
\def\gnl{g_{NL}}
\def\taunl{\tau_{NL}}

\def\bigoh{{\mathcal O}}

\def\bigoh{{\mathcal O}}

\begin{document}

\begin{titlepage}
\setcounter{page}{1} \baselineskip=15.5pt \thispagestyle{empty}

\bigskip\
\begin{center}
{\fontsize{17.5}{30}\selectfont  \bf The Non-Gaussian Halo Mass Function} \\
\vskip 5pt
{\fontsize{17.5}{30}\selectfont \bf with $f_{NL}$, $g_{NL}$ and $\tau_{NL}$}
\end{center}
\vspace{0.5cm}
\begin{center}
{\fontsize{14}{30}\selectfont Marilena LoVerde$^1$ and Kendrick M.~Smith$^2$}
\end{center}

\begin{center}
\vskip 8pt
\textsl{${}^1$ Institute for Advanced Study, Einstein Drive, Princeton, NJ 08540, USA}
\vskip 4pt
\textsl{${}^2$ Princeton University Observatory, Peyton Hall, Ivy Lane, Princeton, NJ 08544 USA}
\end{center}
\vspace{1.2cm}

\hrule \vspace{0.3cm}
{ \noindent \textbf{Abstract} \\[0.2cm]
\noindent
Primordial non-Gaussianity has emerged as one of the most promising probes of the inflationary epoch. 
While the cosmic microwave background and large-scale halo bias currently provide the most stringent 
constraints on the non-Gaussian parameter $\fnl$, the abundance of dark matter halos is a complementary  probe which may allow tests of Gaussianity which are independent of the precise form of non-Gaussian initial conditions.
We study the halo mass function in $N$-body simulations with a range of non-Gaussian initial conditions. 
In addition to the usual $\fnl$ model, we consider $\gnl\Phi^3$-type non-Gaussianity and models where the 
4-point amplitude $\taunl$ is an independent parameter. 
We introduce a new analytic form for the halo mass function in the presence of primordial non-Gaussianity, 
the ``log-Edgeworth'' mass function, and find good agreement with the $N$-body simulations.
The log-Edgeworth mass function introduces no free parameters and can be constructed from first principles for 
any model of primordial non-Gaussianity.
}
 \vspace{0.3cm}
 \hrule

\vspace{0.6cm}
\end{titlepage}

\newpage

\section{Introduction}
\label{sec:intro}
One of the most exciting prospects in observational cosmology is the opportunity to constrain the physics of inflation, thereby probing energy scales which are far beyond the reach of accelerator experiments \cite{Guth:1980zm,Guth:1982ec,Hawking:1982cz,Starobinsky:1982ee,Bardeen:1983qw,Kamionkowski:1996zd,Seljak:1996gy}. Current cosmic microwave background (CMB) data provides strong evidence for a spatially flat universe with small density perturbations drawn from a nearly scale-invariant power spectrum, in accord with inflationary predictions \cite{Komatsu:2010fb}. Nevertheless, distinguishing between microphysical models on the basis of the scalar power spectrum alone remains a challenge. For single-field, slow-roll inflation, higher-order non-Gaussian statistics of the curvature perturbation are unobservably small \cite{Acquaviva:2002ud,Maldacena:2002vr, Creminelli:2004yq}. However, there are broad classes of inflationary models -- those that violate slow-roll, have multiple fields, or modified kinetic terms for example -- that can generate observable levels of non-Gaussianity (see for example, \cite{Bartolo:2004if,Cheung:2007st,Flauger:2010ja, Senatore:2010wk} and references therein). A detection of non-Gaussianity would therefore rule out single-field, slow-roll inflation and could also be a powerful discriminator between these alternative scenarios. 

At present the tightest constraints on non-Gaussianity in $\Phi(\x)$, the primordial curvature perturbation\footnote{In 
Eq.~(\ref{eq:localIC}) and throughout the rest of this paper, we have defined a curvature $\Phi = \frac{3}{5} \zeta$, where
$\zeta$ is the primordial curvature fluctuation conserved on super-horizon scales.  This notation is conventional in studies of primordial non-Gaussianity, although possibly confusing since the Bardeen curvature $\Phi_H$ is equal to $\frac{2}{3} \zeta$ at early times when Eq.~(\ref{eq:localIC}) is applied but $\frac{3}{5} \zeta$ in the matter dominated era, long after the non-Gaussianity in  Eq.~(\ref{eq:localIC}) is imprinted.},  come from constraining the amplitude of several higher-point ``shapes'' inspired by different inflationary scenarios \cite{Komatsu:2003iq,Babich:2004gb,Creminelli:2005hu,Yadav:2007yy,Meerburg:2009ys,Smith:2009jr,Senatore:2009gt}.  For instance, in the so-called local model \cite{Salopek:1990jq,Gangui:1993tt,Komatsu:2001rj,Okamoto:2002ik,Enqvist:2008gk}, the initial curvature is a non-Gaussian field defined through
\be
\label{eq:localIC}
\Phi(\x)=\Phi_G(\x)+\fnl\left(\Phi_G(\x)^2-\langle\Phi_G^2\rangle\right)+\gnl\left(\Phi_G(\x)^3-3\langle\Phi_G^2\rangle\Phi_G(\x)\right)+\dots
\ee
where $\Phi_G$ is a Gaussian field and $\fnl,\gnl$ are free parameters.
The WMAP constraints on these parameters are $-10< f_{NL}<74$ \cite{Komatsu:2010fb} and $-7.4 \times 10^5 < \gnl < 8.2\times 10^5$ \cite{Smidt:2010sv} or $ -12.34\times 10^5 < \gnl <15.58\times 10^5$ \cite{Fergusson:2010gn} at $95\%$ confidence. 
This model also generates a scale-dependent signature in the bias of dark matter halos \cite{Dalal:2007cu,Matarrese:2008nc,McDonald:2008sc,Afshordi:2008ru} that allows for 
competitive constraints from low-redshift data: $-29<\fnl< 70$ \cite{Slosar:2008hx}, and $-3.5\times 10^5<\gnl<8.2 \times 10^5$ \cite{Desjacques:2009jb}
at 95\% CL.
The Planck CMB satellite is expected to achieve $1$-$\sigma$ errors that are smaller by a factor of $3$--$5$ \cite{Komatsu:2001rj,Baumann:2008aq}.

Signatures of primordial non-Gaussianity can also appear the abundance of dark matter halos \cite{Lucchin:1987yv,Colafrancesco:1989px,Chiu:1997xb,Robinson:1998dx,Robinson:1999wh,Koyama:1999fc,Verde:2000vr,Matarrese:2000iz}. For instance, positive (negative) skewness in the density field will tend to increase (decrease) the number of very high mass halos hosting galaxy clusters. The number density of halos as inferred from the cluster mass function has been shown to be a probe of primordial non-Gaussianity which is complementary to the CMB \cite{Sefusatti:2006eu,LoVerde:2007ri,Oguri:2009ui,Roncarelli:2009pp,Sartoris:2010cr,Cunha:2010zz,Fedeli:2010ud} (see also e.g. \cite{Berge:2009xj, Pace:2010sr,Maturi:2011am,Marian:2010mh} for investigations of weak lensing as a probe of primordial non-Gaussianity). The mass function is sensitive to cumulants beyond the $3$-point function but relatively insensitive to the precise shape of the $N$-point functions. Therefore, the mass function can constrain non-Gaussianity without prior knowledge of template shapes, but is less powerful for discriminating between forms of non-Gaussian initial conditions.  At present evidence of {\it any} primordial non-Gaussianity would be extraordinary, so it is useful to obtain observational constraints from a variety of methods. Recently there have been hints of an overabundance of high-$z$ massive clusters \cite{Jimenez:2009us,Holz:2010ck,Hoyle:2010ce,Enqvist:2010bg} which can be interpreted as evidence for primordial non-Gaussianity (but note that \cite{Cayon:2010mq,Mortonson:2010mj,Williamson:2011jz} find consistency with a Gaussian mass function).  This picture will undoubtedly sharpen in the near future with improved mass function constraints from experiments such as SDSS \cite{Hao:2010zze}, Chandra \cite{Vikhlinin:2008ym}, ACT \cite{Marriage:2010cp}, SPT \cite{Vanderlinde:2010eb}, and Planck \cite{Planck:2011aj}. 

In this paper we study the halo mass function in $N$-body simulations with non-Gaussian initial conditions with two forms of non-negligible trispectra: (i) initial conditions where the connected $4$-point function has the same ``shape'' as that from the $\fnl^2$ terms in Eq.~(\ref{eq:localIC}) but boosted in amplitude relative to the $3$-point function (the ``$\tau_{NL}$'' model ) and (ii) initial conditions with the  $\gnl$ contribution from Eq.~(\ref{eq:localIC}) much larger than that from $\fnl$.
This extends current high resolution studies of the halo mass function with pure $\fnl$-type initial conditions \cite{Dalal:2007cu,Kang:2007gs,Grossi:2007ry,Pillepich:2008ka,Grossi:2009an}. See also  \cite{Desjacques:2009jb} for studies of the mass function from $N$-body simulations with a cubic term $\gnl$ in the initial conditions, and $\cite{Wagner:2010me}$ for $N$-body simulations with initial conditions with more general primordial bispectra.

We propose an analytic halo mass function, the ``log-Edgeworth'' mass function, which accurately describes our simulations for a wide range of $\fnl$, $\gnl$, and $\taunl$. 
The log-Edgeworth mass function is conceptually similar to the second-order Edgeworth mass function from \cite{LoVerde:2007ri}, 
but is a better fit to $N$-body simulations in cases where the two disagree, in particular for the high-mass limit, where the Edgeworth mass function breaks down.

Throughout this paper we use the WMAP5+BAO+SN fiducial cosmology \cite{Dunkley:2008ie}:
baryon density $\Omega_bh^2 = 0.0226$, cold dark matter (CDM) density $\Omega_ch^2 = 0.114$, Hubble parameter $h=0.70$,
spectral index $n_s=0.961$, optical depth $\tau = 0.080$, and power-law initial curvature power spectrum 
$k^3 P_\zeta(k) / 2\pi^2 = \Delta_\zeta^2 (k/k_{\rm piv})^{n_s-1}$ where $\Delta_\zeta^2 = 2.42 \times 10^{-9}$
and $k_{\rm piv} = 0.002$ Mpc$^{-1}$.

In \S \ref{sec:introIC} we introduce the generalized local non-Gaussian initial conditions considered in this paper. In \S \ref{sec:mass_function_theory} we discuss prescriptions for analytic mass functions that describe the effects of non-Gaussianity through the cumulants of the density field smoothed on scale $M$, and also present fitting formulae for the smoothed skewness and kurtosis.  The $N$-body simulations and a comparison with analytic mass functions is presented in \S \ref{sec:massfcnsims}.  Concluding remarks are given in \S \ref{sec:discussion}. Appendix \ref{app:cumulantscalc} contains a discussion of calculations of the smoothed skewness, kurtosis and the $\mathcal{O}(\fnl^2)$ correction to the variance. 

\section{Non-Gaussian Initial Conditions}
\label{sec:introIC}

The simplest model of primordial non-Gaussianity is the ``local'' type, in which the initial curvature
$\Phi$ is given by
\be
\Phi(\x) = \Phi_G(\x) + f_{NL} ( \Phi_G(\x)^2 - \langle \Phi_G^2 \rangle )  \label{eq:fnl_def}
\ee
and the 3-point and connected 4-point functions of the initial curvature are given by
\be
\langle \Phi(\k_1) \Phi(\k_2) \Phi(\k_3) \rangle = f_{NL} P_\Phi(k_1) P_\Phi(k_2) (2\pi)^3 \delta_D\left( \sum \k_i \right) + \mbox{(5 perm.)} + \bigoh(\fnl^3)
\ee
\be
\langle \Phi(\k_1) \Phi(\k_2) \Phi(\k_3) \Phi(\k_4) \rangle_c = 2 f_{NL}^2 P_\Phi(k_1) P_\Phi(k_2) P_\Phi(|\k_1+\k_3|) (2\pi)^3 \delta_D\left( \sum \k_i \right) + \mbox{(23 perm.)} + \bigoh(\fnl^4)
\ee
where $\delta_D$ is the Dirac delta function and $P_\Phi$ is the power spectrum of $\Phi_G$. 
This type of non-Gaussianity is generated, e.g.~in the curvaton model, in which there is a second light field
present during inflation (the curvaton) that decays after inflation has ended and generates the primordial
curvature perturbation \cite{Linde:1996gt,Lyth:2001nq,Lyth:2002my,Sasaki:2006kq,Huang:2008zj}. 
We assume throughout that the curvaton decays before dark matter freezout, so that no dark matter isocurvature mode is generated.

In this section, we will describe two sets of non-Gaussian initial conditions which generalize Eq.~(\ref{eq:fnl_def})
and give rise to a trispectrum in the density field that is large enough to affect the halo mass function. 

\subsection{Equal Power from the Curvaton and Inflaton: An Example of $\tau_{NL}\neq \left(\frac{6}{5}\fnl\right)^2$}
\label{ssec:introcurv}
Most studies of the curvaton model have focused on the case where the curvaton completely dominates the primordial curvature perturbation. An alternative set-up, considered recently by \cite{Ichikawa:2008iq,Tseliakhovich:2010kf,Smith:2010gx,Shandera:2010ei}, supposes that the curvaton and inflaton contribute equally to the primordial curvature perturbation. If the inflaton contribution is assumed to be Gaussian and the curvaton contribution is non-Gaussian, the statistics of the initial conditions are changed in important ways.  In particular the higher $N$-point functions are boosted in amplitude relative to the bispectrum, and the boost depends on the ratio of inflaton-to-curvaton contributions to the curvature. 

More precisely, the primordial curvature in this model is given by
\be
\label{eq:zipzc}
\Phi(\x) = \Phi_{i}(\x) + \Phi_c(\x)
\ee
where $\Phi_i$ and $\Phi_c$ denote inflaton and curvaton contributions.
We assume that $\Phi_i$ and $\Phi_c$ are uncorrelated fields with proportional power spectra, i.e.
$P_{\Phi_i}(k) = \frac{\xi^2}{1+\xi^2} P_\Phi(k)$ and $P_{\Phi_c}(k) = \frac{1}{1+\xi^2} P_\Phi(k)$,
where $\xi$ is a free parameter which represents the ratio of inflaton to curvaton contributions.

We take $\Phi_i$ to be Gaussian and $\Phi_c$ to be a field with local non-Gaussianity:
\be
\Phi_c(\x) = \Phi_{c,G}(\x) + \tilde f_{NL} \left(\Phi_{c,G}^2(\x)-\langle \Phi_{c,G}^2(\x)\rangle\right)
\label{eq:fnlcmb_def}
\ee
where $\Phi_{c,G}$ is a Gaussian field.

In this model the power spectrum, bispectrum, and trispectrum of the initial curvature are given by:
\ba
\label{eq:2pt}
\langle \Phi(\k_1) \Phi(\k_2) \rangle &=& P_\Phi(k_1) (2\pi)^3 \delta_D(\k_1+\k_2) + \bigoh(\fnl^2) \\
\label{eq:3pt}
\langle \Phi(\k_1) \Phi(\k_2) \Phi(\k_3) \rangle &=& \fnl (2\pi)^3 \delta_D\left(\sum\k_i \right) P_{\Phi}(k_1) P_{\Phi}(k_2) + \mbox{(5 perm.)} + \bigoh(\fnl^3) \\
\label{eq:4pt}
\langle \Phi(\k_1)\Phi(\k_2)\Phi(\k_3)\Phi(\k_4)\rangle_c &=& \frac{\taunl}{(6/5)^2} 2
(2\pi)^3\delta_D\left(\sum\k_i\right) P_\Phi(k_1) P_\Phi(k_2) P_\Phi(|\k_1+\k_3|) \nn \\
&& \hspace{1cm} + \mbox{(23 perm.)}+\mathcal{O}(\fnl^4)
\label{eq:npoints}
\ea
where we have defined
\ba
f_{NL} &=& \frac{\tilde f_{NL}}{(1+\xi^2)^2} \nn \\
\tau_{NL} &=& \left(\frac{6 f_{NL}}{5}\right)^2 (1+\xi^2) \label{eq:2fderiv}\,.
\ea
The bispectrum and trispectrum in this model have the same shapes as in the curvaton-dominated model considered
previously (Eq.~(\ref{eq:fnl_def})), but the coefficients $f_{NL}$, $\tau_{NL}$ are independent parameters.
The curvaton-dominated model corresponds to the special case $\tau_{NL} = (\frac{6}{5} f_{NL})^2$ and 
the current bounds are $-6000<\tau_{NL}<33,000$ at $95\%$ confidence \cite{Smidt:2010sv}.
The factor $(6/5)$ here is conventional and has been introduced for consistency with the literature.
Our perspective is that primordial non-Gaussianity is most conveniently parameterized
by the coefficients of the primordial $N$-point functions, and the specific two-field model in this
section is just a mechanism for generating local non-Gaussianity with prescribed $f_{NL}$ and $\tau_{NL}$.
For this reason,
we will use $f_{NL}$ and $\tau_{NL}$ as the basic parameters of the model,
and treat $\tilde f_{NL}$ and $\xi$ as derived parameters.

\subsection{Local Initial Conditions with Kurtosis but No Skewness: $g_{NL}$ }
\label{ssec:gnl}
Another variation on initial conditions with local non-Gaussianity is to consider a case where the quadratic term in Eq.~(\ref{eq:localIC}) vanishes but the cubic term is included \cite{Okamoto:2002ik}. Initial conditions of this form can be generated in a curvaton model where the potential for the curvaton has terms that are not quadratic and cancellations from these terms set $\fnl \sim 0$ while generating a large $\gnl$ \cite{Enqvist:2005pg,Sasaki:2006kq,Enqvist:2008gk,Huang:2008bg}
\be
\label{eq:gNLIC}
\Phi(\x)=\Phi_{G}(\x) + \gnl\left(\Phi_G^3(\x)-3 \langle\Phi^2_G\rangle\Phi_G(\x)\right)\,.
\ee
We have found it convenient to include the $-3 \langle\Phi^2_G\rangle \Phi_G(\x)$ term in the definition
of the $\gnl$ model so that the power spectrum of $\Phi$ will be unchanged to first order in $\gnl$.\footnote{An 
alternate convention (e.g.~\cite{Desjacques:2009jb}) omits this term from the definition;
in this case the ``bare'' power spectrum amplitude $\Delta_\zeta^2$ used to set up the Gaussian
field $\Phi_G$ will differ from the observed value of $\Delta_\zeta^2$ that would be
inferred by measuring observable power spectra.
The two definitions will be equivalent in any analysis which marginalizes $\Delta_\zeta^2$, but
the first requires less bookkeeping to keep track of the difference between bare and observable
power spectrum amplitudes.}
Note that the expectation value $\langle \Phi^2_G \rangle$ is infrared divergent, but converges in a finite volume.
We compute it as a discrete sum over Fourier modes $\k \ne 0$ in our simulation, for consistency with the way the Gaussian
initial conditions are generated.

With initial conditions given by~(\ref{eq:gNLIC}), the power spectrum and connected trispectrum are given by:
\ba
\langle \Phi(\k_1)\Phi(\k_2)\rangle&=&(2\pi)^3\delta_D(\k_1+\k_2)P_{\Phi}(k_1)+\mathcal{O}(\gnl^2)\\
\langle \Phi(\k_1)\Phi(\k_2)\Phi(\k_3)\Phi(\k_4)\rangle_{\textrm{conn.}}&=& g_{NL}(2\pi)^3\delta_D\left(\sum\k_i \right) P_\Phi(k_1)P_\Phi(k_2)P_\Phi(k_3) \nn \\ && \hspace{1cm} + \textrm{(23 perm.)} + \mathcal{O}(\gnl^2)  \label{eq:gnl_trispectrum}
\ea
and all odd $N$-point functions vanish.

\section{Halo Mass Functions - Theory}
\label{sec:mass_function_theory}

There are a number of non-Gaussian mass functions in the literature. Most of them are derived from some expansion of a non-Gaussian probability distribution 
function (PDF) for the mass density field and either follow the Press-Schechter ansatz \cite{Press:1973iz,Chiu:1997xb,Robinson:1999se,Matarrese:2000iz,LoVerde:2007ri,Chongchitnan:2010xz}, 
or more formally use excursion set theory \cite{Lam:2009nb,Maggiore:2009rx,DeSimone:2010mu,D'Amico:2010ta}. 
These mass functions are typically specified by the PDF or $N$-point correlation functions of the linear density field (or equivalently the initial curvature).
On the other hand,  Pillepich, Porciani and Hahn \cite{Pillepich:2008ka}  present a 
fitting formula for the mass function derived from non-Gaussian simulations which (in addition to the usual cosmological parameters)
depends on $f_{NL}$, assuming $g_{NL}=0$ and $\tau_{NL} = (\frac{6}{5} f_{NL})^2$.
These mass functions are all in relatively good agreement with each other and with simulations with
$\fnl$-type (i.e. Eq.~(\ref{eq:fnl_def})) non-Gaussian initial conditions \cite{Pillepich:2008ka,Grossi:2009an}. 
In this paper, our emphasis will be on analytic mass functions that depend only on the cumulants of the variable $\delta_M$,
the linearly-evolved density fluctuation smoothed on mass scale $M$ \cite{LoVerde:2007ri}. 
However, see \cite{Maggiore:2009hp,D'Amico:2010ta,DeSimone:2010mu} for extensions based on excursion set theory 
which include additional terms from so-called unequal time correlators. 

In this section we describe a general formalism for deriving non-Gaussian mass functions, and apply it to the case of local non-Gaussianity
parameterized by $\fnl$, $\gnl$, and $\taunl$.
Our approach is conceptually similar to the Edgeworth approach from \cite{LoVerde:2007ri} but differs in some details which we now explain.
We'll follow the Press-Schechter model \cite{Press:1973iz} which states that the fraction $F(M)$ of volume collapsed to objects of mass $\ge M$ is equal
to the probability for $\delta_M$ to exceed the collapse threshold $\delta_c\approx 1.42$.\footnote{Throughout this paper
we will include a correction to the spherical collapse threshold $\delta_c=1.686$, $\delta_c\rightarrow\delta_c\sqrt{q}$ with $q=1/\sqrt{2}$
which has been shown to give better agreements with simulations \cite{Pillepich:2008ka,Grossi:2009an}.}
In the presence of primordial non-Gaussianity, the 1-point PDF $\rho(\delta_M)$ is perturbed, and this leads to a change in the mass function
which is computable in the Press-Schechter model.

The Edgeworth expansion is a representation of a general PDF as a power series in the higher cumulants of the distribution.
Since higher cumulants parameterize deviations from Gaussianity, the Edgeworth expansion is most useful in the regime of
weak non-Gaussianity, where the series converges rapidly.
In \cite{LoVerde:2007ri}, the Edgeworth expansion for $\rho(\delta_M)$ was truncated
to obtain an estimate of the $\fnl$ dependence of $\rho(\delta_M)$, from which the $\fnl$ dependence of the mass function
can be calculated.
Here, we will find it convenient to truncate the series expansion for the quantity $\ln(F(M))$ (rather than the quantity $\rho(\delta_M)$);
we call the non-Gaussian mass function obtained in this way the ``log-Edgeworth'' mass function.
We will find that the log-Edgeworth mass function is a better fit to simulations than the Edgeworth expansion for parameter
values where the two disagree, in particular for the high-mass limit, where the Edgeworth expansion breaks down.

Another more minor detail is that we keep a few more terms in the series expansion than are usually quoted from \cite{LoVerde:2007ri}. We do this 
for the following reason: the simulations in \S\ref{sec:massfcnsims} will show that the mass function is $\tau_{NL}$-dependent, in the case where 
$\taunl$ is varied at fixed $\fnl$ (and with $\gnl=0$).
Therefore, to allow $\taunl$ dependence, we will keep the term in the Edgeworth series which is first order in $\taunl$.
Since $\taunl$ is the same order as $\fnl^2$, we will also keep terms of order $\fnl^2$ in order to truncate the series
consistently.
Similarly, we keep terms of first order in $\gnl$ but not second order, since $\gnl$ has the same order as $\taunl$ when it appears
in $N$-point correlation functions.

\subsection{Cumulants}
\label{ssec:cumulants}

The linear density field smoothed on scale $M$ is given by
\be
\label{eq:smoothedd}
\delta_M(z)=\int\frac{d^3\k}{(2\pi)^3} W_{M}(k) \alpha(k,z) \Phi(\k)
\ee
where
\be
\label{eq:W}
W_M(k)=\frac{3\sin(kR(M))}{(kR(M))^3}-\frac{3\cos(kR(M))}{(kR(M))^2}
\ee
is the Fourier transform of a tophat window whose comoving radius $R(M) = (3M/4\pi\rho_m)^{1/3}$ encloses mass $M$, and
\be
\alpha(k,z) = \frac{2 k^2 T(k) D(z)}{3 \Omega_m H_0^2}
\ee
is defined so that $\delta(\k,z) =  \alpha(k,z) \Phi(\k)$ in linear theory. Here, $T(k)$ denotes the matter transfer function, and $D(z)$ is the linear growth function normalized to
$D(z) = 1/(1+z)$ at high $z$.

The reduced cumulants are defined as
\be
\label{eq:cumulants}
\kappa_N(M)=\frac{\langle \delta_M^N\rangle_{\rm conn.}}{\sigma(M)^N} \qquad \textrm{for $N\ge 3$}
\ee
where $\sigma^2(M)=\langle \delta_M^2\rangle$. 
Note that $\kappa_N(M)$ is independent of $z$ as implied by the notation. 

For the non-Gaussian mass function in the next subsection, we will need to know the
cumulants $\kappa_3(M)$ and $\kappa_4(M)$ to leading order in the non-Gaussianity
parameters $\fnl$, $\gnl$ and $\taunl$.
These cumulants can be calculated either by Monte Carlo, or analytically by integrating the $N$-point
correlation function (Eqs.~(\ref{eq:3pt}),~(\ref{eq:4pt}),~(\ref{eq:gnl_trispectrum})) over
the wavenumbers $\k_i$ with appropriate weighting.
The reduced cumulants (with the $\sigma(M)^N$ denominator) are slowly varying functions of $M$ and 
relatively insensitive to the assumed cosmological parameters but slightly cumbersome to compute.
Details of the calculation are given in Appendix~\ref{app:cumulantscalc}.
Here, we simply quote the results: for $\kappa_3$, we find the following to be a good fitting function
\be
\label{eq:k3fit}
\kappa_3(M) \approx \fnl \, (6.6 \times 10^{-4}) \left(1- 0.016 \ln\left(\frac{M}{h^{-1}M_\odot}\right)\right)
\ee
(see also \cite{Chongchitnan:2010xz}). The cumulant $\kappa_4$ is more subtle: it contains terms proportional to $\gnl$ and $\taunl$,
and the $\bigoh(\taunl)$ term formally diverges as the volume of the simulation box is taken to infinity.
This divergence is well-known and we discuss it in detail in Appendix~\ref{app:cumulantscalc}.
For now, we just quote a fitting function for $\kappa_4$ with explicit dependence on the box size $L$:\footnote{In
Eqs.~(\ref{eq:k4fit}) and~(\ref{eq:k2fit}), the
factor $\Delta_\Phi^2 \ln(L/L_0)$ is actually an approximation to the exact infrared divergent behavior which assumes
$(1-n_s)\ln(L/L_0) \ll 1$.  If this condition is not satisifed then a more accurate approximation can be obtained by
making the replacement
\be
\Delta_\Phi^2 \ln\left(\frac{L}{L_0}\right) \rightarrow \left( \frac{k^2 P_\Phi(k)}{2\pi^2} \right)_{k=4.67/L} 
  \frac{(L/L_0)^{n_s-1}-1}{n_s-1} \nn
\ee
}
\ba
\label{eq:k4fit}
\kappa_4(M) & \approx & \gnl \, (1.6 \times 10^{-7}) \left(1 - 0.021 \ln \left(\frac{M}{h^{-1}M_\odot}\right)\right) \nn \\
&& \hspace{0.5cm} + \frac{\taunl}{(6/5)^2} \, \left[ (6.9\times 10^{-7}) \left(1- 0.021 \ln\left(\frac{M}{h^{-1}M_\odot}\right)\right)
  + 48 \Delta_\Phi^2 \ln\left( \frac{L}{L_0} \right) \right]
\ea
where $L_0 = 1600$ $h^{-1}$ Mpc and $\Delta_\Phi^2 = \frac{9}{25} \Delta_\zeta^2$ is equal to $(8.72\times 10^{-10})$ for the
fiducial cosmology from \S\ref{sec:intro}.

We will also need the leading non-Gaussian contribution to the variance $\sigma(M)^2$, which has the same order\footnote{Strictly speaking, the amplitude of $\kappa_2$ need not be proportional to $\tau_{NL}^2/\fnl^2$ but may be a free parameter. The form of $\kappa_2$ given in Eq.~(\ref{eq:k2fit}) assumes initial conditions with $\fnl$ and $\taunl$ which are implemented using the two-field model from \S\ref{ssec:introcurv}.} (i.e.~quadratic in $\fnl$ in the case $\taunl=(6/5\fnl)^2$) as the second term in $\kappa_4$.  We write the variance as a sum of a Gaussian term $\sigma_G^2(M)$ and a non-Gaussian term
$\sigma_G^2(M) \kappa_2(M)$, where $\kappa_2$ represents the fractional correction due to primordial non-Gaussianity
\be
\label{eq:k2def}
\sigma^2(M) \equiv \sigma_G^2(M) \left(1 + \kappa_2(M) \right)
\ee
and find the following fitting function for $\kappa_2(M)$
\be
\label{eq:k2fit}
\kappa_2(M) \approx \frac{\taunl^2}{(6/5)^4 \fnl^2} 
  \left[ (4.0\times 10^{-8}) \, \left( 1 - 0.021 \ln\left(\frac{M}{h^{-1} M_\odot}\right)\right)
    + 4 \Delta_\Phi^2 \ln\left( \frac{L}{L_0} \right) \right] 
\ee
where $L_0 = 1600$ $h^{-1}$ Mpc.  Note that $\kappa_2(M)$ is also infrared divergent.

The correct choice of $L$ in the infrared divergent parts of the above cumulants depends on the context in which a non-Gaussian mass
function is desired.
When comparing to $N$-body results in \S\ref{sec:massfcnsims}, we will choose $L$ to be the side length of the (periodic) simulation volume.
When comparing to the observed mass function from real data, a reasonable choice would be $L \sim 2R$ where $R\approx 14000$ Mpc is
the comoving causal horizon.
The need to choose a value of $L$ may seem unfortunate, but it is a generic feature of models with local non-Gaussianity that has
nothing to do with mass functions {\em per se}: such models only make sense when regulated in a finite volume, and physical
quantities (such as the non-Gaussian contribution to the power spectrum) depend weakly on this choice.

\begin{figure}[t]
\begin{center}
\epsfxsize=8cm\epsffile{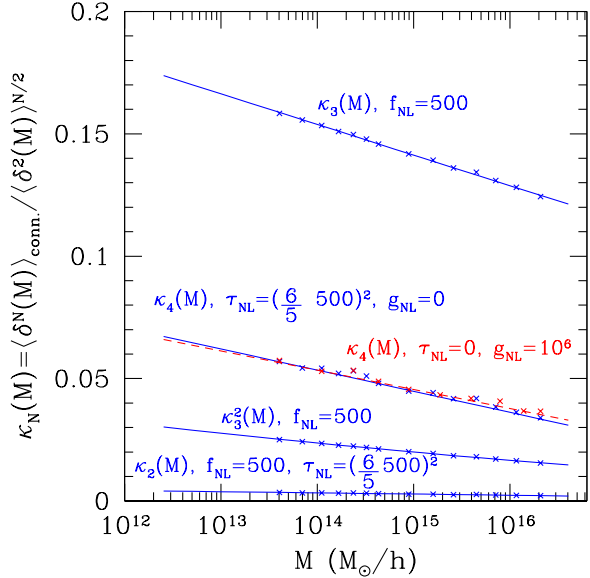}
\end{center}\caption{The reduced cumulants of the density field smoothed on scale $M$. The points are the cumulants measured from Monte-Carlo realizations of the initial conditions, the lines are the fits from Eqs.~(\ref{eq:k3fit})--(\ref{eq:k2fit}). Note that the smoothed kurtosis has roughly the same $M$ dependence for the $\gnl$ (dashed, red line) and $\tau_{NL}$ (solid, blue line) models despite the different ``shapes'' of their trispectra. For $\kappa_2$ and $\kappa_{4,\tau_{NL}}$ we have assume $L=1600\, h^{-1}$Mpc. }
\label{fig:cumulants}
\end{figure}

In Fig.~\ref{fig:cumulants}, we plot the reduced cumulants $\kappa_2$, $\kappa_3$, $\kappa_4$ obtained by Monte Carlo,
with the fitting functions shown for comparison.

\subsection{Mass Function Derivation}
\label{ssec:mfcnderive}

In the Press-Schechter model, the fraction of volume $F(M)$ collapsed to halos of mass $M$ is given by
\be
F(M) = \int_{\nu_c(M)}^\infty d\nu \, \rho(\nu,M)  \label{eq:Fdef}
\ee
where $\rho(\nu,M)$ is the 1-point PDF of the variable $\nu=\delta_M/\sigma(M)$, 
and $\nu_c(M)=\delta_c/\sigma(M)$ with $\delta_c=1.42$.
The mass function $n(M)$ is then given by
\be
n(M) = -2\frac{\rho_m}{M} F'(M)  \label{eq:ndef}
\ee
where primes denote derivatives with respect to $M$.

In a Gaussian cosmology, we have $\rho(\nu,M) = (2\pi)^{-1/2} \exp(-\nu^2/2)$.
In a non-Gaussian cosmology, we can expand the PDF as a series (the Edgeworth expansion):
\be
\rho(\nu,M) = \frac{\exp{(-\nu^2/2)}}{(2\pi)^{1/2}} (1 + p_1(\nu,M) + p_2(\nu,M) + \cdots )
\ee
where
\ba
p_1(\nu,M) &=& \frac{1}{6} \kappa_3(M) H_3(\nu)  \\
p_2(\nu,M) &=& \frac{1}{2} \kappa_2(M) H_2(\nu) + \frac{1}{24} \kappa_4(M) H_4(\nu) + \frac{1}{72} \kappa_3(M)^2 H_6(\nu)
\ea
and $H_n(\nu)$ are the Hermite polynomials defined by $H_n(\nu) = (-1)^n e^{\nu^2/2} \frac{d^n}{d\nu^n}e^{-\nu^2/2}$. 
Note that $p_1$ represents contributions which are
first order in $\fnl$, and $p_2$ represents contributions which are either second order in $\fnl$
or first order in $\gnl$, $\taunl$. It is worth noting that the explicit mass dependence of $p_1$, $p_2$, arising from the mass 
dependence of the $\kappa(M)$'s (as well as the box dependence in $\kappa_2(M,L)$, $\kappa_4(M,L)$) 
breaks the universality of the mass function. However as can be seen in Eqs.~(\ref{eq:k4fit}),~(\ref{eq:k2fit}) and Fig.~\ref{fig:cumulants}, the box dependent contributions are small
and the mass dependence is slight, so universality is only weakly broken. 

Plugging into Eqs.~(\ref{eq:Fdef}),~(\ref{eq:ndef}), we can obtain series expansions for $F(M)$ or $n(M)$.
The series $F(M) = F_0(M) + F_1(M) + F_2(M)$ is given by
\ba
F_0(M)&=& \frac{1}{2} \mbox{erfc} \left( \frac{\nu_c(M)}{\sqrt{2}} \right) \label{eq:F0} \\
F_1(M) &=& \frac{1}{(2\pi)^{1/2}} e^{-\nu_c(M)^2/2} \left(\frac{ \kappa_3(M)}{6} H_2(\nu_c(M)) \right) \\
F_2(M) &=& \frac{1}{(2\pi)^{1/2}} e^{-\nu_c(M)^2/2} \bigg( 
  \frac{\kappa_2(M)}{2} H_1(\nu_c(M)) + \nn \\
&& \hspace{1cm} + \frac{\kappa_4(M)}{24} H_3(\nu_c(M)) + \frac{\kappa_3(M)^2}{72} H_5(\nu_c(M)) \bigg)\label{eq:F2def}\,.
\ea
It will also be convenient to have expressions for the derivatives with respect to $M$
\ba
F_0'(M) &=& -\frac{\nu_c'(M)}{(2\pi)^{1/2}} e^{-\nu_c(M)^2/2} \\
F_1'(M) &=& F_0'(M)\left(\frac{\kappa_3(M)}{6} H_3(\nu_c(M)) -\frac{\kappa_3'(M)}{6\nu_c'} H_2(\nu_c(M)) \right) \\
F_2'(M) &=& F_0'(M)\left( \frac{\kappa_2(M)}{2} H_2(\nu_c(M)) + \frac{\kappa_4(M)}{24} H_4(\nu_c(M)) + \frac{\kappa_3(M)^2}{72} H_6(\nu_c(M)) \right. \nn \\
&& \left. - \frac{\kappa'_2(M)}{2\nu_c'} H_1(\nu_c(M)) - \frac{\kappa_4'(M)}{24\nu_c'} H_3(\nu_c(M))- \frac{\kappa_3(M) \kappa_3'(M)}{36\nu_c'} H_5(\nu_c(M)) \right) \label{eq:F2prime}\,.
\ea
The ``Edgeworth'' mass function from \cite{LoVerde:2007ri} is defined by truncating the series for $F(M)$ (or for $n(M)$).
At second order in $\fnl$ (first order in $\gnl$, $\tau_{NL}$) the resulting mass function is
\ba
\label{eq:edge2}
\left.\frac{n_{NG}}{n_G}\right|_{\textrm{Edgeworth}}\approx \left(1+\frac{F'_1(M)}{F'_0(M)} + \frac{F'_2(M)}{F'_0(M)}\right)\,.
\ea
This mass function has been shown to be in good agreement with simulations for $\tau_{NL}=(\frac{6}{5}\fnl)^2$ \cite{Pillepich:2008ka,Grossi:2009an}. 
However, it has a couple of slightly annoying properties that arise at large $|\fnl|$; if truncated at first order in $\fnl$ it is unbounded from below 
as $\fnl$ approaches large negative values, if truncated at second order the mass function is non-monatonic at 
the high-mass end for negative $\fnl$.\footnote{We have experimented with using a mass-dependent barrier $\nu_c(M)$ in Eqs.~(\ref{eq:F0})-(\ref{eq:F2def}), 
 chosen so that the Gaussian mass function derived in this way agrees with the Sheth-Tormen or Warren mass functions
  \cite{Sheth:2001dp,Warren:2005ey} but found it did not improve the Edgeworth mass function.}

Inspired by these issues, we propose a slightly different mass function. 
If we truncate the series for $\ln(F(M))$ rather than $F(M)$:
\be
\ln(F(M)) \approx \ln F_0(M) + \frac{F_1(M)}{F_0(M)} + \frac{F_2(M)}{F_0(M)} - \frac{1}{2} \left( \frac{F_1(M)}{F_0(M)} \right)^2  \label{eq:logF_truncated}
\ee
then we obtain the mass function
\ba
\label{eq:modedge2}
\left.\frac{n_{NG}}{n_G}\right|_{\textrm{log-Edgeworth}}&\approx& \exp\left[ \frac{F_1(M)}{F_0(M)} + \frac{F_2(M)}{F_0(M)} - \frac{1}{2} \left( \frac{F_1(M)}{F_0(M)} \right)^2 \right] \\
  && \times \left( 1 + \frac{F_1'(M) + F_2'(M)}{F_0'(M)} - \frac{F_1(M) F_1'(M)}{F_0(M) F_0'(M)}
                     - \frac{F_1(M) + F_2(M)}{F_0(M)} + \frac{F_1(M)^2}{F_0(M)^2} \right)\,.\nn
\ea
We will refer to this non-Gaussian mass function as the ``log-Edgeworth'' mass function. 

The Edgeworth~(\ref{eq:edge2}) and log-Edgeworth~(\ref{eq:modedge2}) mass functions agree to second order in $\fnl$
(or first order in $\gnl$, $\taunl$) but differ in the way the low-order derivatives are extrapolated to finite values
of $\fnl$, $\gnl$ and $\taunl$. 
As we will see in the next section, the log-Edgeworth mass function is a better fit to $N$-body simulations in the
regime where the two disagree, and also provides sensible asymptotic behavior in the limit of large halo mass.
In principle, if we kept all terms in each series expansion the Edgeworth and log-Edgeworth mass functions would agree.
It is interesting to note, that at lowest order in $\fnl$, $\gnl$, the ratio $n_{NG}/n_G$ as predicted by the log-Edgeworth mass function reduces to that from the Edgeworth mass function in 
the limit $\nu_c(M)<<1$ and the ``MVJ" \cite{Matarrese:2000iz} mass function for $\nu_c(M)>>1$.\footnote{We are grateful to Vincent Desjacques for pointing this out.}

Note that we have written both the Edgeworth and log-Edgeworth mass functions as expressions for $n_{NG}/n_G$, to separate
the issue of analytically describing the Gaussian mass function $n_G$ 
(see e.g.~\cite{Jenkins:2000bv,Warren:2005ey,Tinker:2008ff})
from the issue of describing the fractional correction due to primordial non-Gaussianity.

In this section, we have presented the log-Edgeworth mass function in maximum generality in order to provide a
 framework which can be adapted to models of non-Gaussianity not considered in this paper. 
If all that is desired is to compute the mass function for specified values of $\fnl$, $\gnl$ and $\taunl$, this
can be done straightforwardly by using the fitting functions in Eqs.~(\ref{eq:k3fit})--(\ref{eq:k2fit}) to compute $\kappa_i(M)$,
then using Eqs.~(\ref{eq:F0})--(\ref{eq:F2prime}) to compute $F_i(M)$ and $F_i'(M)$,
and finally using Eq.~(\ref{eq:modedge2}) to compute $n_{NG}/n_G$.

\section{Halo Mass Function from $N$-body Simulations}

\label{sec:massfcnsims}
\begin{figure}[!ht]
$\begin{array}{cc}
\epsfxsize=8cm\epsffile{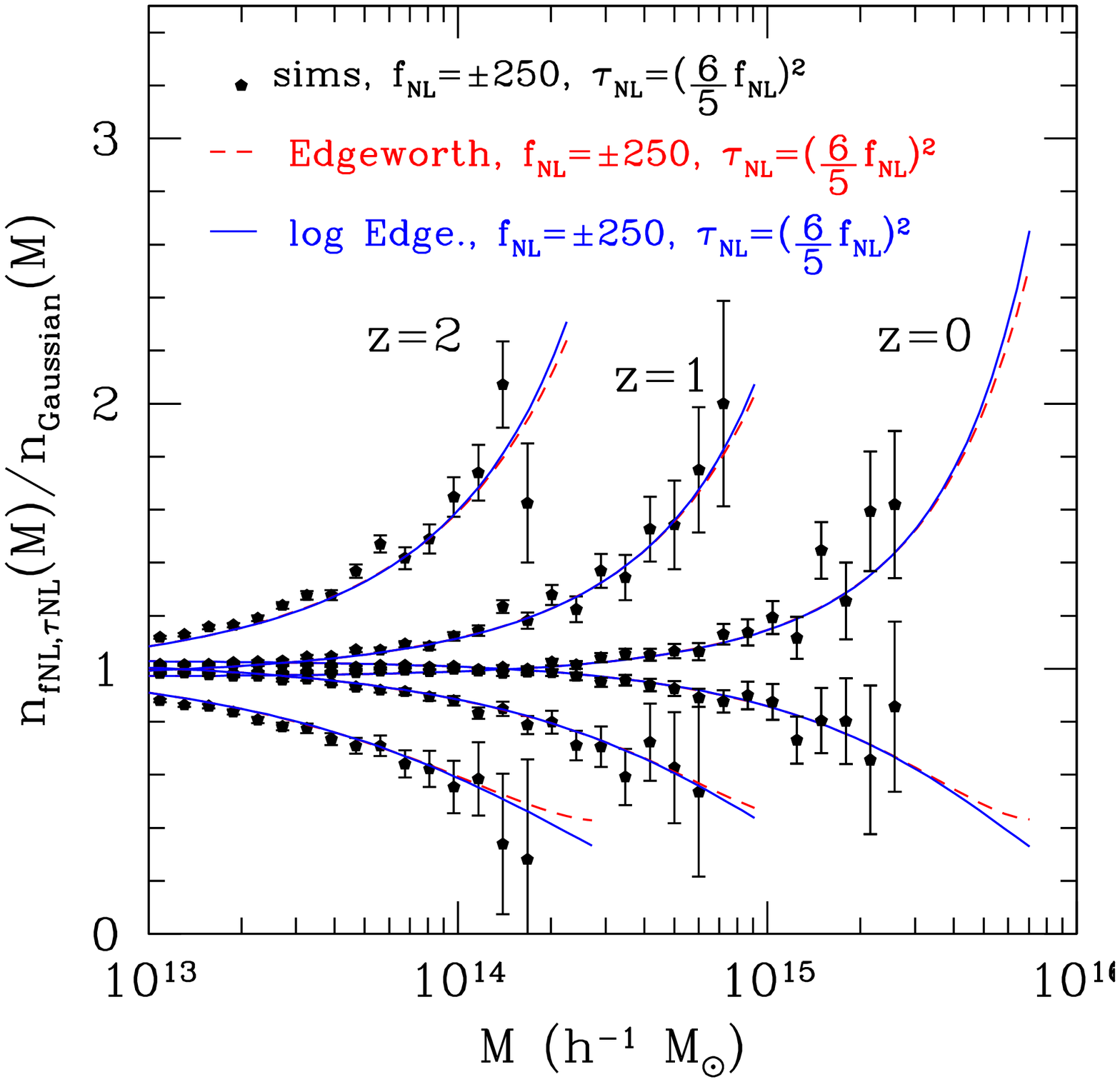}& \epsfxsize=8cm\epsffile{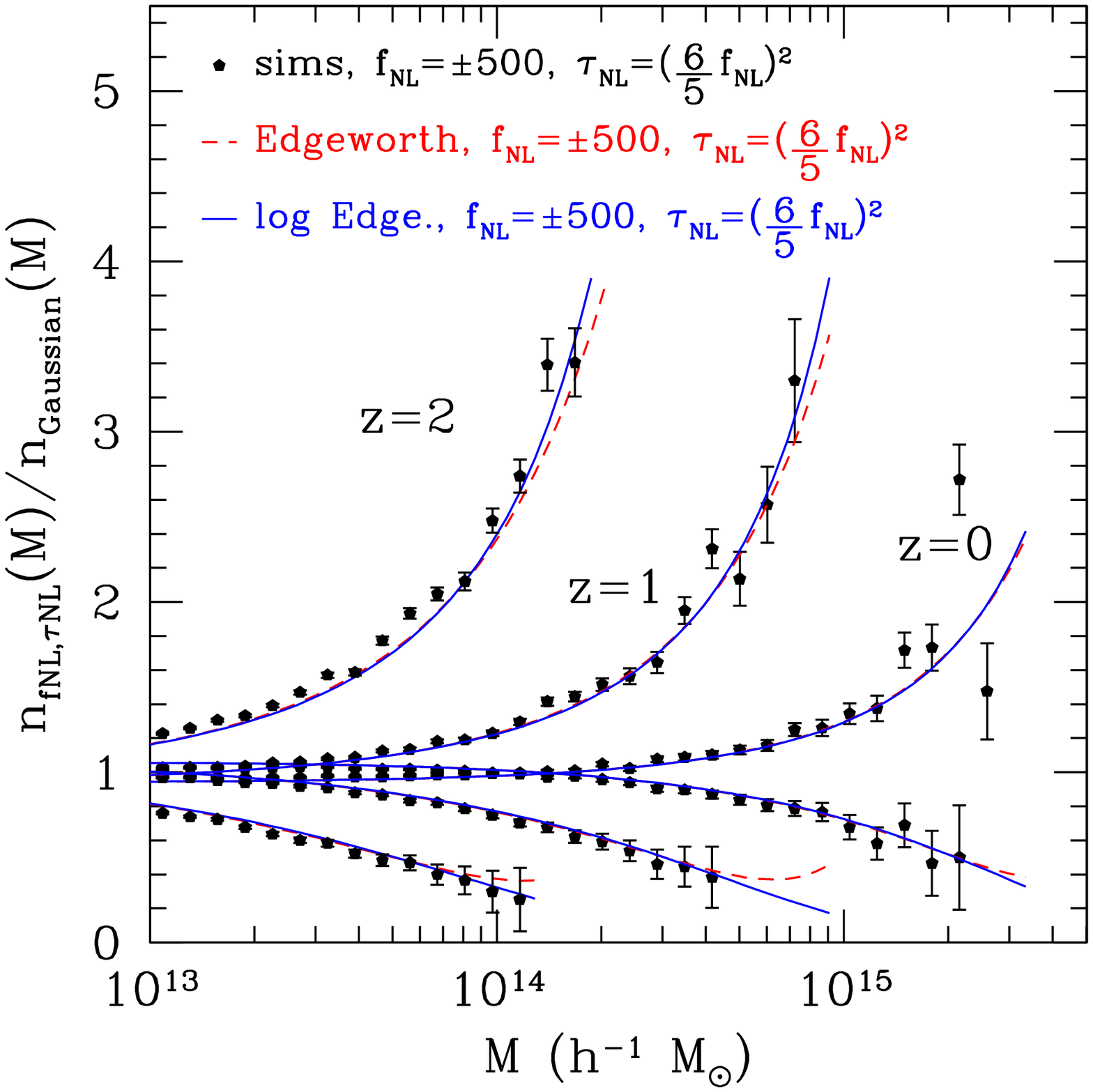}\\
\mbox{(a)} & \mbox{(b)} \\
%
%
\epsfxsize=8cm\epsffile[18 144 592 760]{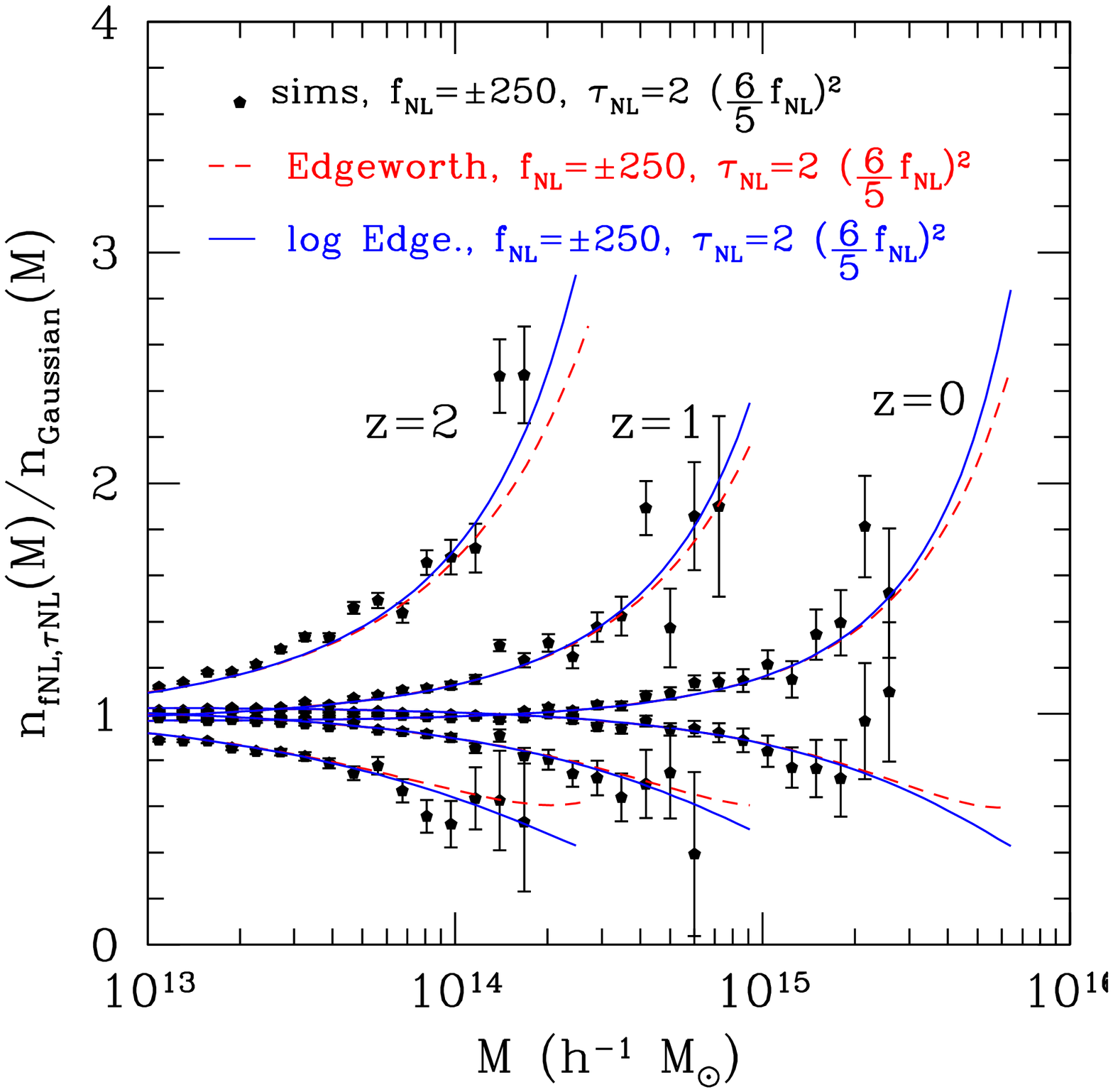}& \epsfxsize=8cm\epsffile[18 144 592 760]{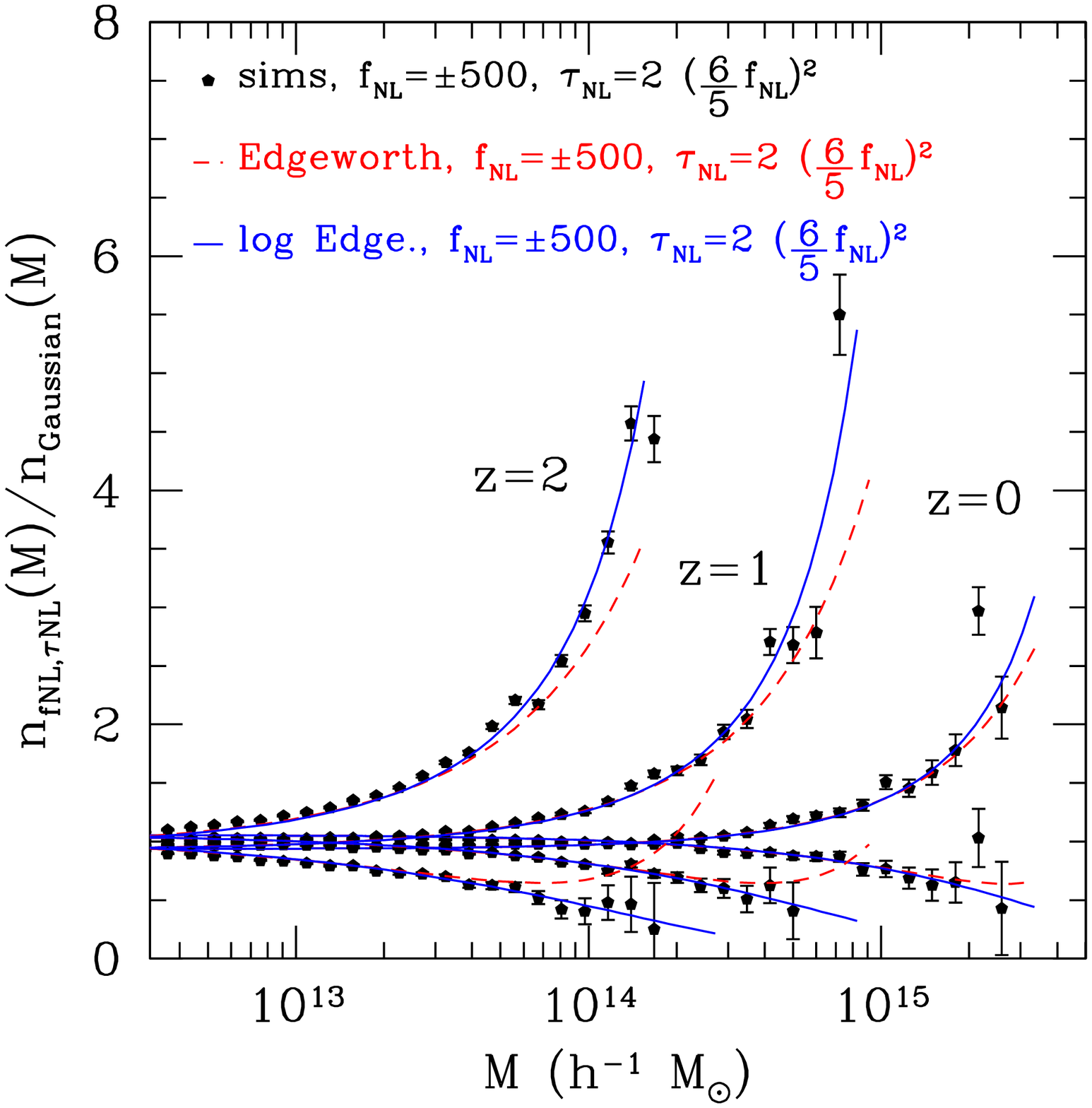}\\
\mbox{(c)} & \mbox{(d)} 
\end{array}$\caption{Comparison of the Edgeworth (Eq.~(\ref{eq:edge2})) and log-Edgeworth (Eq.~(\ref{eq:modedge2})) mass functions
for non-Gaussian initial conditions with nonzero $\fnl$ and $\taunl$.
For $\taunl=(\frac{6}{5}\fnl)^2$ (i.e.~perturbations generated entirely by the curvaton) they both provide reasonably good fits. 
For $\taunl=2(\frac{6}{5}\fnl)^2$ (i.e.~equal power from the curvaton and inflaton) the log-Edgeworth mass function is in better agreement.}
\label{fig:mass_functioncompare}
\end{figure}

\begin{figure}[!ht]
$\begin{array}{cc}
\epsfxsize=8cm\epsffile{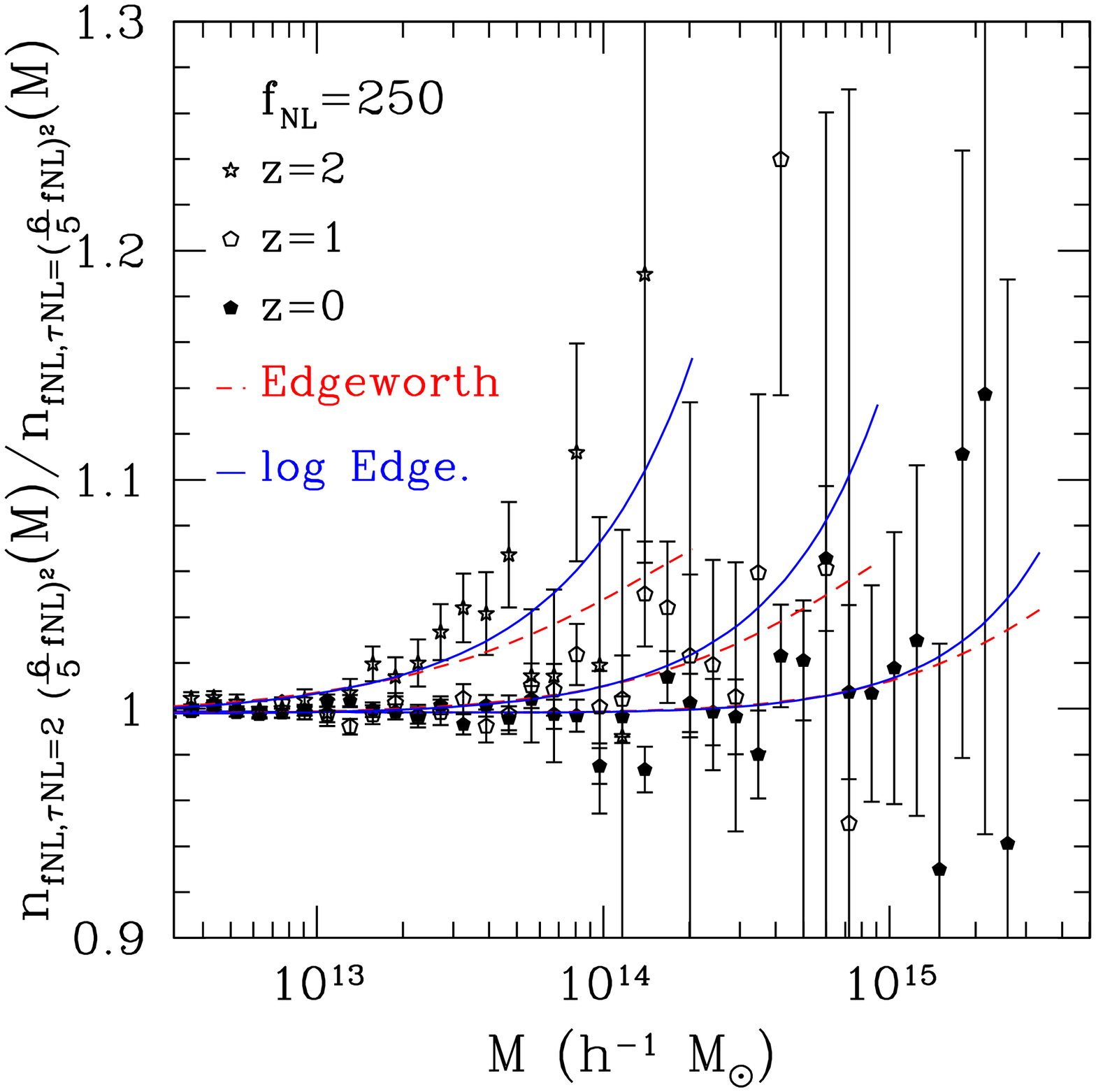}& \epsfxsize=8cm\epsffile{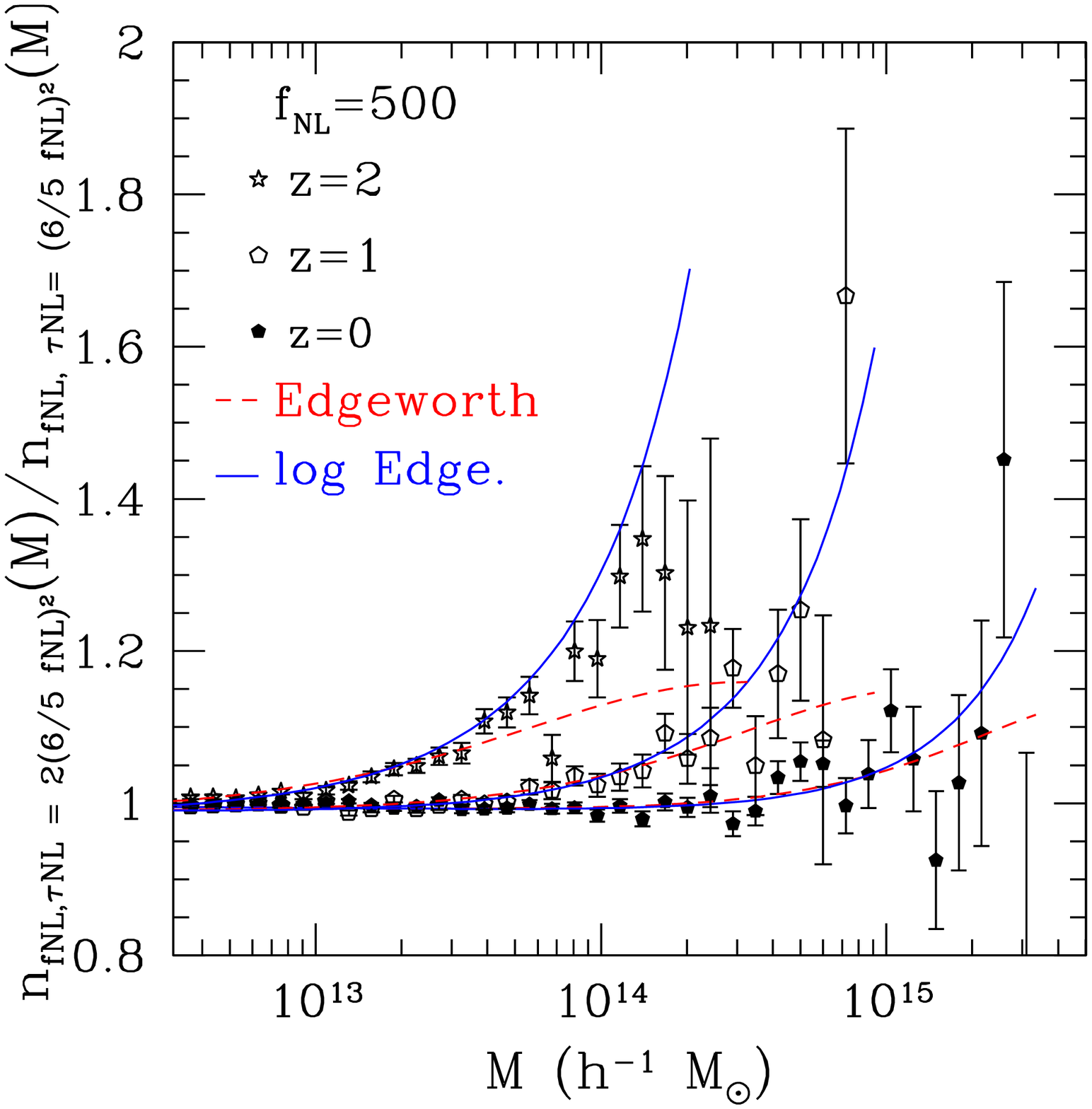}\\
\mbox{(a)} & \mbox{(b)} \\
\epsfxsize=8cm\epsffile[18 144 592 760]{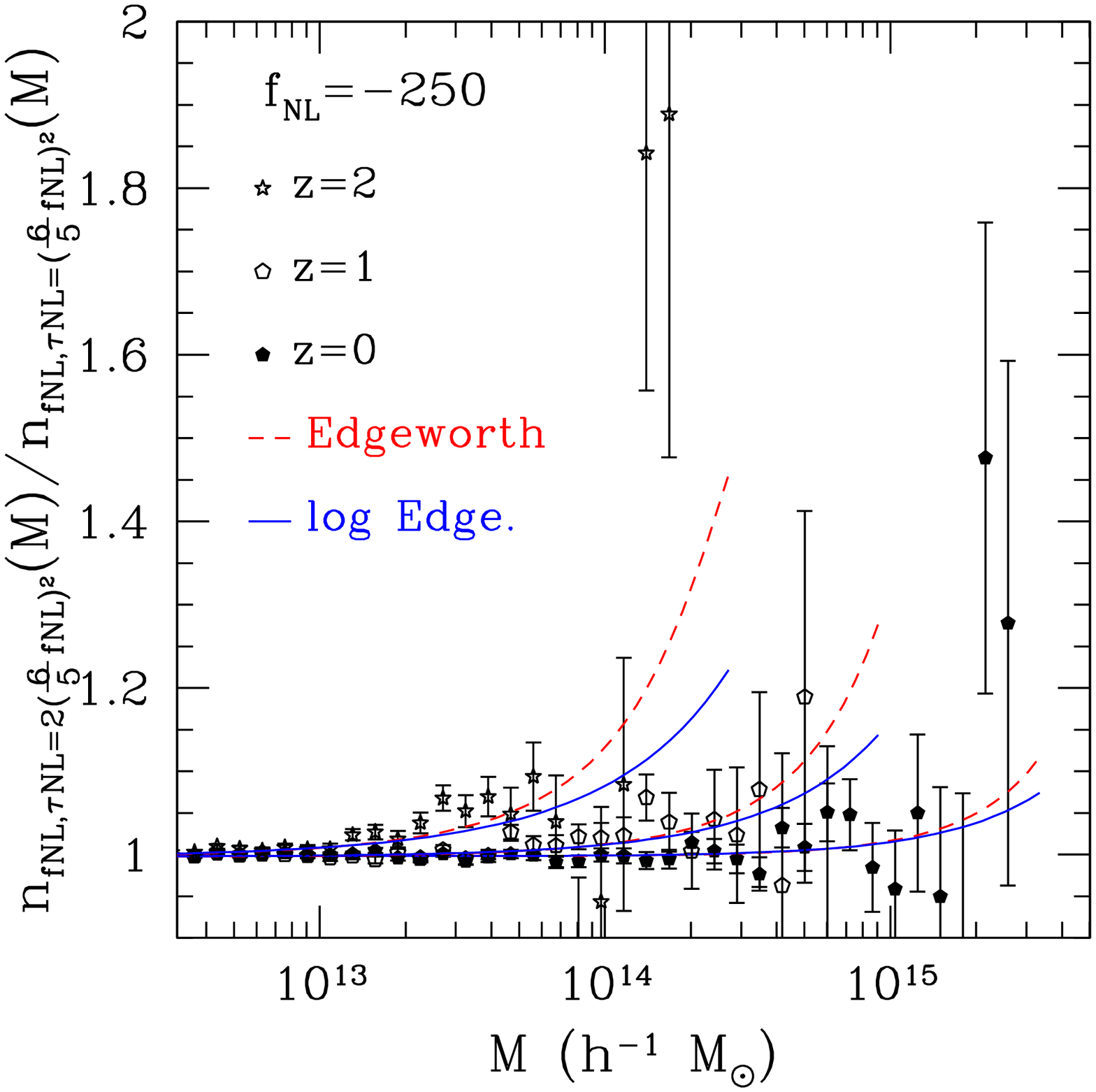}& \epsfxsize=8cm\epsffile[18 144 592 760]{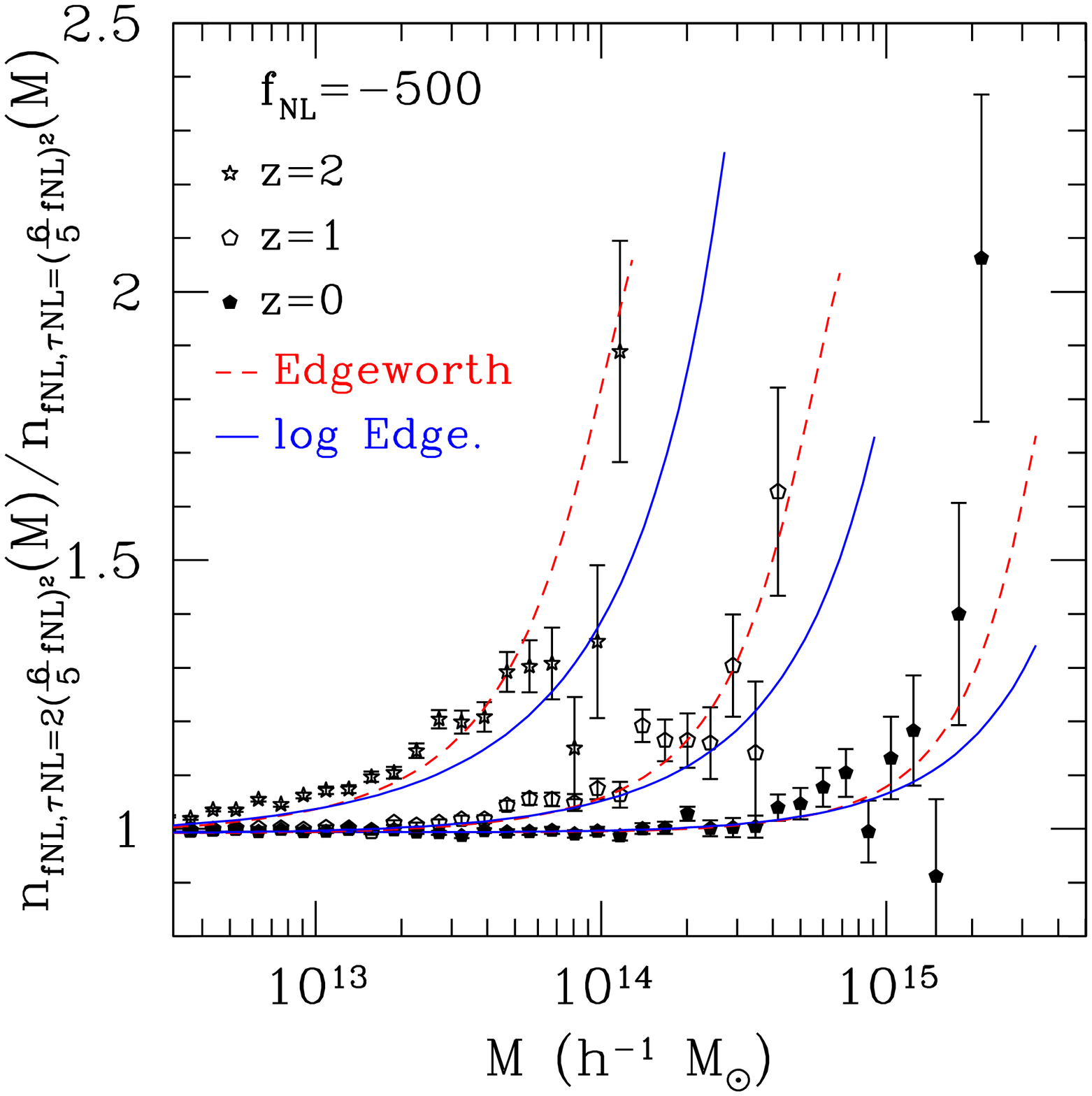}\\
\mbox{(c)} & \mbox{(d)} 
\end{array}$\caption{Comparison of mass functions with $\taunl = (\frac{6}{5}\fnl)^2$ and $\taunl = 2(\frac{6}{5}\fnl)^2$
for: (a)$\fnl=+250$ (b)$\fnl=+500$ (c)$\fnl=-250$ and (d) $\fnl=-500$.
This plot is intended to isolate the effect of the primordial trispectrum on the halo mass function, since the two
models being compared have equal bispectra but different trispectra.
The  curves are the ``Edgeworth'' mass function in Eq.~(\protect{\ref{eq:edge2}}) and the ``log-Edgeworth'' mass function presented in  Eq.~(\protect{\ref{eq:modedge2}}).}
\label{fig:mass_functioncomparex1x0}
\end{figure}

\begin{figure}[!ht]
$\begin{array}{cc}
\epsfxsize=8cm\epsffile{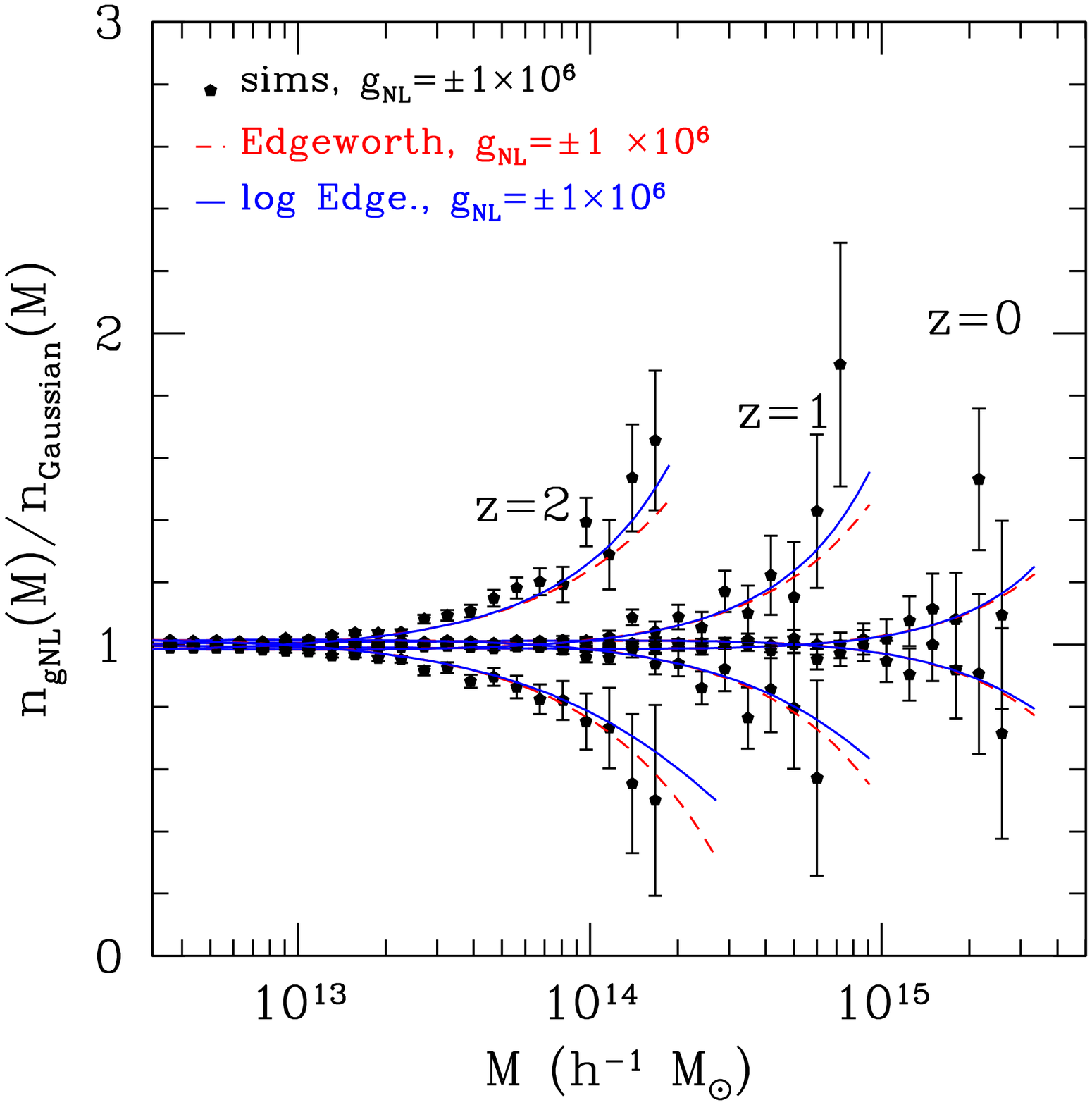}&\epsfxsize=8cm\epsffile{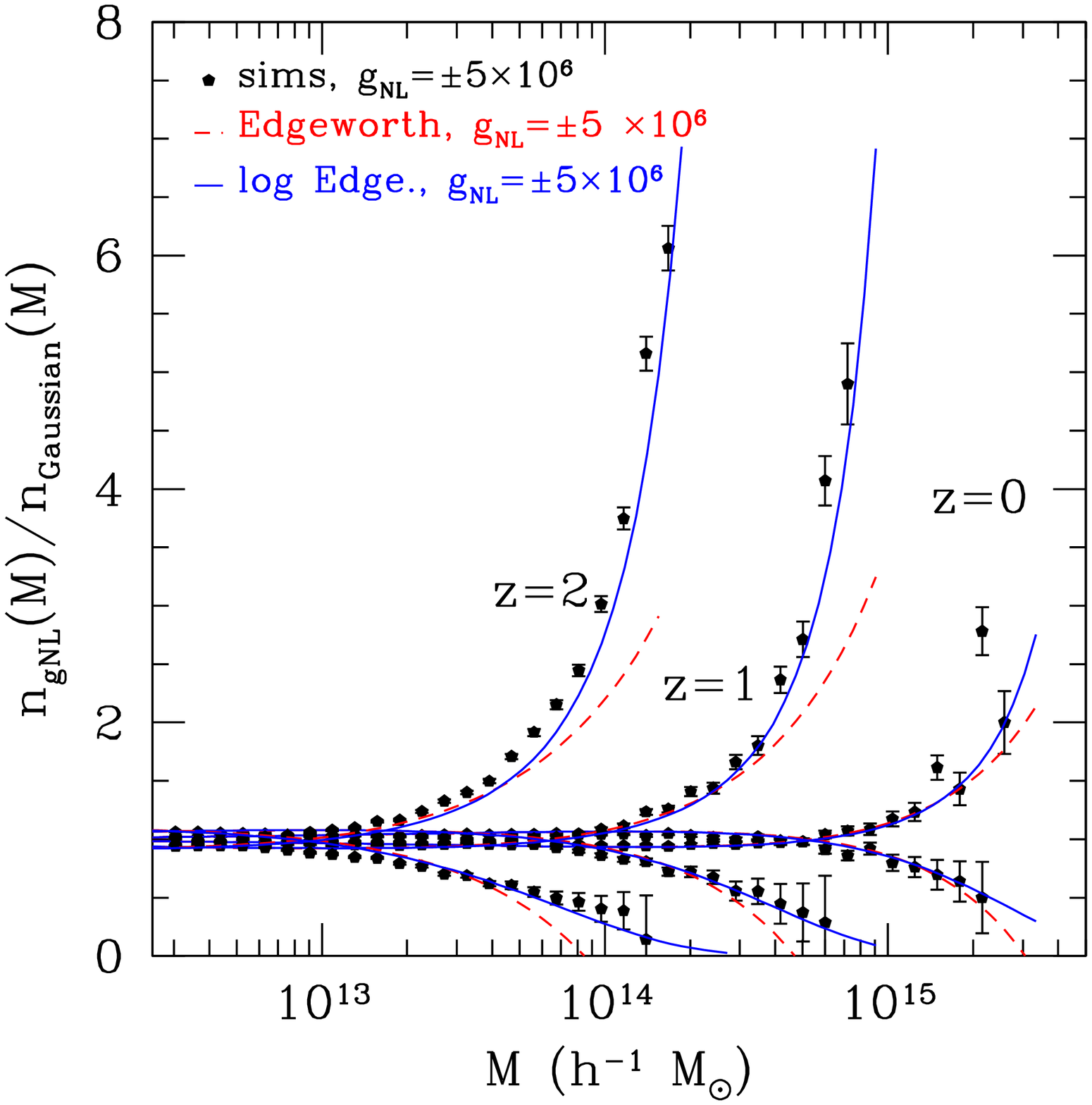}\\
\mbox{(a)} & \mbox{(b)}
\end{array}$
\caption{Comparison of the Edgeworth (Eq.~(\ref{eq:edge2})) and log-Edgeworth (Eq.~(\ref{eq:modedge2})) mass functions
for initial conditions with a $\gnl\Phi^3$ term (Eq.~(\ref{eq:gNLIC})) (here $\fnl=\tau_{NL}=0$).}
\label{fig:gNLcompare}
\end{figure}

To study the mass function, we
performed collisionless $N$-body simulations using the GADGET-2 TreePM code \cite{Springel:2005mi}.
Simulations were done using periodic
box size $R_{\rm box} = 1600$ $h^{-1}$~Mpc, particle count $N_p = 1024^3$, and force softening
length $R_s = 0.05 (R_{\rm box}/N_p^{1/3})$.
With these parameters and the fiducial cosmology from \S\ref{sec:intro}, the particle mass 
is $m_p = 2.92 \times 10^{11}$ $h^{-1}$~$M_\odot$.

We make non-Gaussian simulations of the initial curvature $\Phi$ in two cases: either (1)
taking $\gnl=0$ and nonzero $\fnl$, $\taunl$, or (2) taking $\fnl=\taunl=0$ and nonzero $\gnl$.
In the first case, we use the two-field model from \S\ref{ssec:introcurv}.
Starting from input parameters $\fnl$ and $\taunl$, we obtain $\tilde{f}_{NL}$ and $\xi$ using Eq.~(\ref{eq:2fderiv}),
then simulate Gaussian fields $\Phi_i$ and $\Phi_{c,G}$ with power spectra $\frac{\xi^2}{1+\xi^2} P_\Phi$
and $\frac{1}{1+\xi^2} P_\Phi$ respectively.
The total initial curvature is then given by $\Phi=\Phi_i+\Phi_c$, where $\Phi_c$ is given by Eq.~(\ref{eq:fnlcmb_def}).
In the second case ($\fnl=\taunl=0$ with $\gnl\ne 0$), we simulate a non-Gaussian $\Phi$
using Eq.~(\ref{eq:gNLIC}).
We do not generate initial conditions with nonzero values for all three parameters $\fnl$, $\gnl$, $\taunl$,
although it would be easy to extend the two-field model to apply in this generality.

Given a realization of the initial curvature $\Phi$, initial particle positions and velocities are generated as follows.
First, we apply the transfer function $T(k)$, computed using CAMB \cite{Lewis:1999bs}, to obtain the Newtonian
potential at the initial redshift $z_{\rm ini}=100$ of the simulations.
Then we obtain initial particle positions using the Zeldovich approximation \cite{Zeldovich:1969sb}. (At $z_{\rm ini}=100$, transient effects due to use of this approximation should be negligible \cite{Crocce:2006ve}.)

After running the $N$-body simulation, we group particles into halos using an MPI parallelized implementation of the 
friends-of-friends algorithm \cite{Frenk:1988zz}
with link length $L_{\rm FOF} = 0.2 R_{\rm box} N_p^{-1/3}$.
For a halo containing $N_{\rm FOF}$ particles, we assign a halo position given by the mean of the individual
particle positions, and a halo mass given by
\be
m_h = m_p \left( N_{\rm FOF} - N_{\rm FOF}^{0.4} \right)\,,
\ee
the second term is recommended in \cite{Warren:2005ey} to minimize particle resolution artifacts in the 
mass function.\footnote{It has been suggested that halos identified by spherical overdensity are better suited to observations 
(although which is the most appropriate halo-finder likely depends on the observable)
and there is a scatter between FOF and spherical overdensity identified halos \cite{White:2002at,Tinker:2008ff}. However, uncertainties in the mass
function from baryonic effects appear to be comparable to uncertainties from halo finding \cite{Lukic:2007fc,Stanek:2008am}.}

In Fig.~\ref{fig:mass_functioncompare} we study the mass function $n(M)$ for a few choices of $(\fnl,\taunl)$. It is seen that the mass function mainly depends on $\fnl$, but some $\taunl$-dependence can also be seen, particularly at high redshift. Recall from Eq.~(\ref{eq:3pt}) that the 3-point function is independent of $\taunl$, so the $\taunl$-dependence at high redshift indicates that the skewness of the density field (in addition to the linear power spectrum) is insufficient to describe the non-Gaussian mass function.  However, the four-point function (Eq.~(\ref{eq:npoints})) and the $\mathcal{O}(\taunl)$ correction to the power spectrum (Eq.~(\ref{eq:k2fit})) do depend on $\taunl$, so our simulations indicate that these higher cumulants are relevant for determining the abundance of high-mass halos. 

The Edgeworth and log-Edgeworth mass functions given in  Eq.~(\ref{eq:edge2})  and Eq.~(\ref{eq:modedge2}) include the first $\taunl$-dependent terms in the series expansion. These mass functions are plotted in Fig.~\ref{fig:mass_functioncompare}.  Both the Edgeworth and log-Edgeworth mass functions appear to provide a good fit to simulations in all but the most extreme cases; only for $\fnl=\pm500$ do we see disagreement. The log-Edgeworth mass function appears to do a better job at describing cosmologies with $\fnl<0$ and also those with $\taunl \ne (\frac{6}{5} \fnl)^2$. 

In Fig.~\ref{fig:mass_functioncomparex1x0} we show the ratio of the non-Gaussian mass function with $\taunl = 2(\frac{6}{5}\fnl)^2$ to the one with $\taunl = (\frac{6}{5}\fnl)^2$, for a fixed value of $\fnl$.  We find that these two mass functions differ by as much as a factor 2 at high redshift and halo mass.
Again both the Edgeworth and log-Edgeworth mass functions do a reasonable job of predicting the effect of varying $\taunl$, though the error bars are rather large and there seems to be some disagreement towards the high mass end for $\fnl=500$. 

Figure~\ref{fig:gNLcompare} shows the non-Gaussian correction to the mass function measured from simulations with $\fnl=\tau_{NL}=0$ and $\gnl=\pm 10^6$ or $\gnl=\pm 5\times 10^6$. As expected, positive $\gnl$ increases the abundance of massive halos while negative $\gnl$ decreases their abundance.  For $\gnl=10^6$, both the Edgeworth and log-Edgeworth mass functions are in reasonable agreement with simulations, but with $\gnl=5\times 10^6$, the log-Edgeworth mass function is in much better agreement.\footnote{There is a caveat: for $\gnl=5\times 10^6$, we find that $\kappa_6(M) \gsim \kappa_4(M)$, which suggests that the $\bigoh(\gnl)$ terms in the series expansion~(\ref{eq:logF_truncated}) are not larger than the $\bigoh(\gnl^2)$ terms which have been neglected.  Nevertheless, we find empirically that the log-Edgeworth mass function~(\ref{eq:modedge2}) agrees with simulations for these values of $\gnl$.  Our perspective is that while this point is somewhat unsettling, it is the simulations rather than the Press-Schechter model that validate the analytic mass function, so the log-Edgeworth mass function does apply for these $\gnl$ values.  The current constraints on $\gnl$ are anyway strict enough to avoid this issue.}

As can be seen in Fig.~\ref{fig:mass_functioncompare} and \ref{fig:gNLcompare}, the improvement from the log-Edgeworth truncation is most 
significant at high masses and redshifts and/or for large non-Gaussianity. For a $3\times 10^{15}$ $h^{-1} M_\odot$ cluster
the difference is $\lsim 10 \%$ at $z=0$ even for $\fnl=\pm 500$ (and $\tau_{NL}=(6/5 \fnl)^2$). On the other hand, if  
$\tau_{NL}=2(6/5\fnl)^2$, the two mass functions differ by ~$10-20\%$ at the same redshift. For more modest 
values of $\fnl$ (~$\pm 100$) there is not a significant difference between the two mass functions for
 $M< 3\times 10^{15}$ $h^{-1} M_\odot$ until $z\gsim 1$ unless $\tau_{NL} \gsim 3(\fnl)^2$.

\begin{figure}[!ht]
$\begin{array}{cc}
\epsfxsize=8cm\epsffile{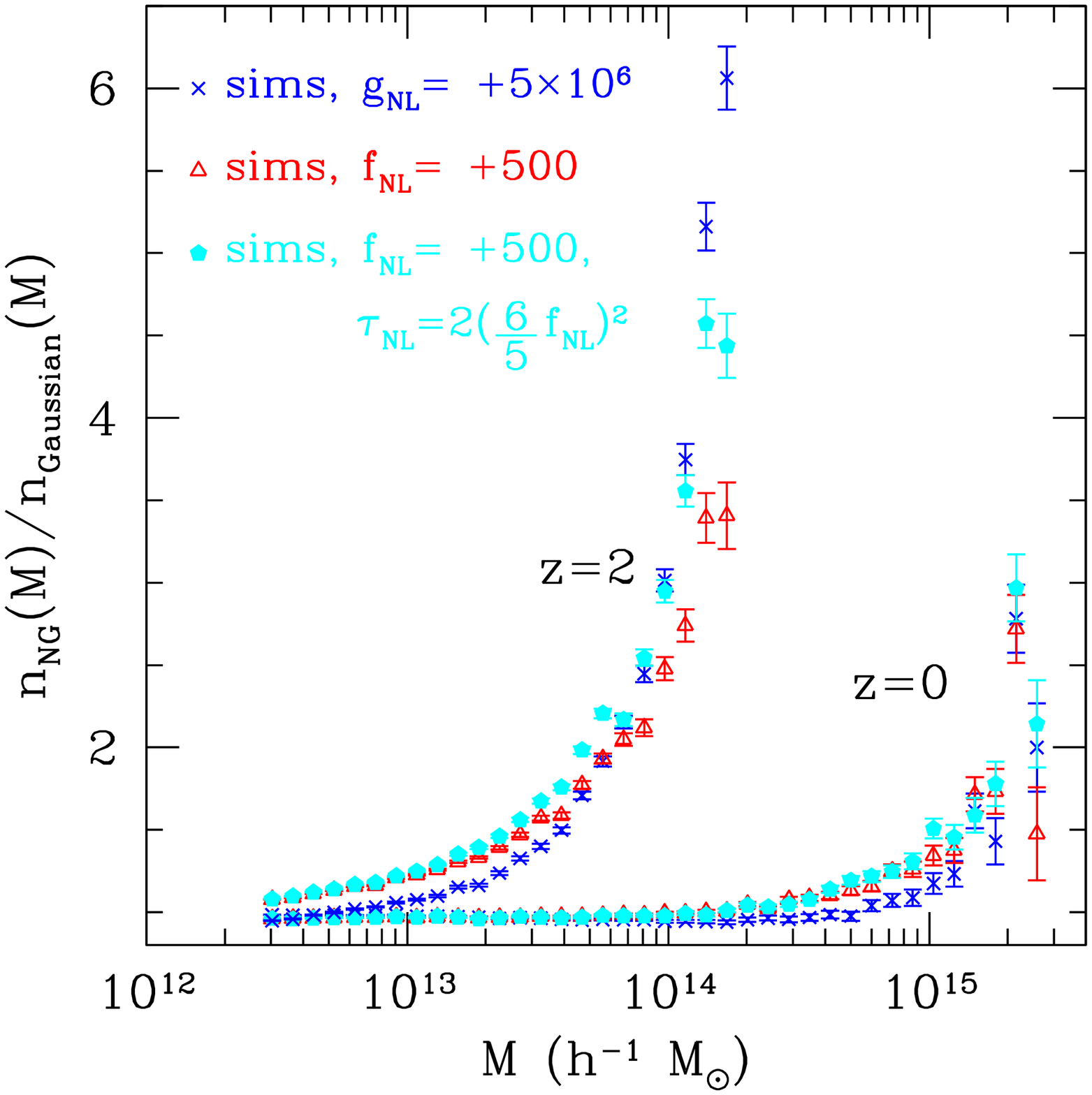}&\epsfxsize=8cm\epsffile{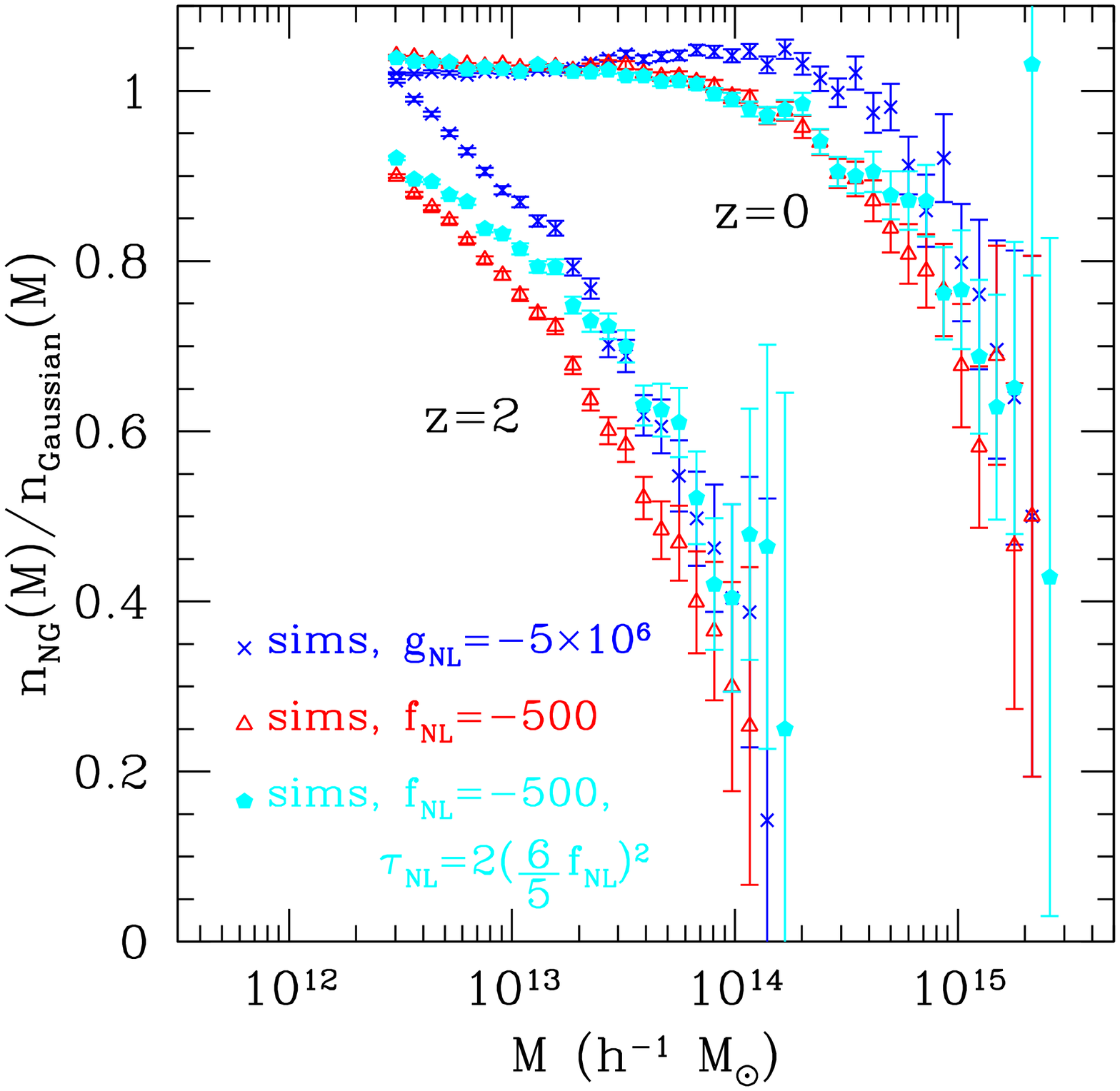}\\
\mbox{(a)} & \mbox{(b)}
\end{array}$
\caption{Comparison of the non-Gaussian corrections to the mass function for initial conditions with $\fnl=\pm 500$ and $\tau_{NL}=(\frac{6}{5}\fnl)^2$; $\fnl=\pm  500$ and $\tau_{NL}=2 (\frac{6}{5}\fnl)^2$; and $\fnl=\tau_{NL}=0$, $\gnl=\pm 5\times 10^6$. The $\tau_{NL}$ models are rather degenerate; a model with $\tau_{NL}\neq (\frac{6}{5}\fnl)^2$ can be made to look like a model with $\tau_{NL}=(\frac{6}{5}\fnl)^2$ by adjusting $\fnl$. On the other hand, $\gnl$ and $\fnl$ change the shape of $n_{NG}/n_G$ in distinct ways. }
\label{fig:gNLcomparefNLgNLtNL}
\end{figure}

Finally, in Fig.~\ref{fig:gNLcomparefNLgNLtNL} we compare the effects of $\fnl$, $\gnl$ and $\tau_{NL}$ on the halo mass function. The shape of  $n_{NG}/n_G$ for the two simulated values of $\tau_{NL}$ is similar. Indeed, from the analytic mass function in Eq.~(\ref{eq:modedge2}) we find that a model with $\tau_{NL}> (\frac{6}{5}\fnl)^2$ can be made to look like a model with $\tau_{NL}=(\frac{6}{5}\fnl)^2$ at most masses by increasing $\fnl$.  On the other hand, $\gnl$ and $\fnl$ change the shape of the mass function in distinct ways:  relative to the $\fnl$ models, $\gnl$ causes a only slow increase/decrease in the abundance of low mass halos but dramatically changes the abundance of very massive halos. This suggests that at least in principle, the halo mass function can distinguish non-Gaussian initial conditions with skewness and kurtosis (i.e.~$\gnl=0$ with nonzero $\fnl$ and $\taunl$) from those with kurtosis only (i.e.~$\fnl=\taunl=0$ with nonzero $\gnl$). 
This is complementary to constraints from large-scale halo clustering, where $\fnl$ and $\gnl$ produce roughly degenerate 
effects \cite{Desjacques:2009jb}, but $\taunl$ can be constrained independently of $\fnl$ by measuring stochasticity in 
the bias \cite{Tseliakhovich:2010kf,Smith:2010gx}.

\section{Discussion}
\label{sec:discussion}

In this paper, we have compared semianalytic predictions for the halo mass function to $N$-body simulations with
``generalized local'' non-Gaussianity parameterized by $\fnl$, $\gnl$ and $\taunl$  (described in \S \ref{sec:introIC}).

Our main result is the log-Edgeworth mass function, Eq.~(\ref{eq:modedge2}), which directly relates the non-Gaussian part of the
halo mass function to non-Gaussianity in the primordial curvature, by expressing $(n_{NG}/n_G)$ in terms of the non-Gaussian
cumulants $\kappa_2$, $\kappa_3$ and $\kappa_4$ (defined in Eqs.~(\ref{eq:cumulants}) and~(\ref{eq:k2def})).
The log-Edgeworth mass function contains no free parameters and is based on the Press-Schechter model for the halo mass function,
expanded as a power series in cumulants.
A strength of this approach, in comparison to a pure fitting function for the non-Gaussian mass function, is that the
log-Edgeworth mass function can be applied to models of primordial non-Gaussianity not explicitly considered in this paper,
since the cumulants can be computed from first principles in a given model.

We have considered the mass function in more detail for two types of non-Gaussian initial conditions.
The first case (\S\ref{ssec:introcurv}) is a ``$\Phi^2$-type'' local model in which the coefficients of the three-point
and four-point functions are independent parameters $\fnl$, $\taunl$.
This is implemented by taking the initial curvature to be a sum of Gaussian and non-Gaussian fields with constant relative amplitude,
as discussed recently in \cite{Tseliakhovich:2010kf}.
The second case (\S\ref{ssec:gnl}) is a ``$\Phi^3$-type'' model with a $\gnl$-term in the initial curvature.

In both of these models, we calculate the cumulants $\kappa_2$, $\kappa_3$, $\kappa_4$ using methods
developed in Appendix~\ref{app:cumulantscalc}, and give fitting functions in Eqs.~(\ref{eq:k3fit})--(\ref{eq:k2fit}).
Thus, for generalized local models with parameters $\fnl$, $\gnl$, $\taunl$,
a completely explicit expression for the non-Gaussian mass function is obtained
by plugging these fitting functions for the $\kappa$'s into Eqs.~(\ref{eq:F0})--(\ref{eq:F2prime})
for the $F$ functions and their derivatives, and then into the log-Edgeworth mass function Eq.~(\ref{eq:modedge2}). 
While we focus on the abundance of dark matter halos at $z\lsim 2$, these expressions may also be useful for other probes of non-Gaussianity that are sensitive to the one-point PDF of the density field, e.g.~the onset of reionization, the abundance of voids or the Lyman-alpha flux PDF \cite{Crociani:2008dt,Viel:2008jj,Kamionkowski:2008sr}.

In the first model, the cumulants are logarithmically divergent at second order in $\fnl$ (or first order in $\taunl$).
This divergence is a real property of $\fnl$ models: if we set up initial conditions via $\Phi = \Phi_G + \fnl (\Phi_G^2 - \langle \Phi_G \rangle^2)$
in a finite box, measurable quantities such as the power spectrum or halo mass function will ``run'' with the box size
and slowly diverge in the infinite-volume limit.
Our interpretation is that $\fnl$ models are only defined if a box size is also specified, and our calculation of
the cumulants includes box size dependence explicitly (which is propagated to box size dependence in the log-Edgeworth
mass function).

Keeping track of the box size dependence is potentially a strength of our approach, since the statistics of an $\fnl$
model in a Hubble-sized box (the relevant scale for observational constraints) may differ from the smaller box
sizes typically used to study $\fnl$ models using simulations.
Using the log-Edgeworth mass function, we can predict the ratio $n(M)_{R_1}/n(M)_{R_0}$ of mass functions between
box sizes $R_1 = 2\times 10^4$ $h^{-1}$ Mpc and $R_0 = 1600$ $h^{-1}$ Mpc.  For $\fnl$ within the current bounds and 
$|\tau_{NL}| \lsim (few) \fnl^2$, we predict that $n(M)_{R_1}$ and $n(M)_{R_0}$ differ by $\lsim 1\%$ for halo masses $ \lsim 3\times 10^{15}  h^{-1}M_\odot$ at $z=0$. 
However, even for low values of $\fnl$ we predict an important difference if $\tau_{NL}$ is allowed to be significantly different from $(\frac{6}{5}\fnl)^2$.
 For example,  if $\fnl=20$ and $\tau_{NL} \sim 10^4$ we find $n(M)_{R_1}/n(M)_{R_0}\sim1.1$ by $M\sim 3\times 10^{14}$ and larger at higher masses. 
 We have not attempted to compare our predictions for the box size dependence with simulations in this paper, leaving this
for future work.

For the first model, we find excellent agreement between the log-Edgeworth mass function and $N$-body simulations for
$\fnl\in\{\pm 250, \pm 500 \}$ and $\taunl\in\{ (\frac{6}{5}\fnl)^2,\, 2(\frac{6}{5}\fnl)^2 \}$.
For the second model, we find excellent agreement for $\gnl\in\{ \pm 1\times 10^6, \pm 5 \times 10^6 \}$
(see Figs.~\ref{fig:mass_functioncompare}--\ref{fig:gNLcompare}).
These parameter ranges are larger than current observational limits and therefore the log-Edgeworth mass function
appears to provide an excellent fit over the observationally relevant range.

The log-Edgeworth mass function constructed in this paper is conceptually similar to the Edgeworth mass function
from \cite{LoVerde:2007ri}; the main difference is that we expand the quantity $\ln(F(M))$ defined in Eq.~(\ref{eq:Fdef})
as a power series in cumulants, rather than the mass function $n(M)$.
The log-Edgeworth and Edgeworth mass functions agree in the limit where many terms in the series expansion are
retained (or the limit of small $\fnl$) but differ in practice if the expansions are truncated at a fixed finite order.
The main difference is that the Edgeworth mass function has non-physical asymptotic behavior at large $M$, but in the observationally relevant range the
log-Edgeworth mass function continues to behave correctly in this limit.
Additionally, even away from the large-$M$ limit, we find that the log-Edgeworth mass function is a slightly better fit to the
simulations than the Edgeworth mass function.
Recent hints of an overabundance of very massive clusters have renewed interest in understanding the effects of 
primordial non-Gaussianity on the abundance of rare objects. Accurately predicting the high-mass, high redshift 
end of the halo mass function is critical to interpreting the significance of systems. The log-Edgeworth mass
function appears to describe the halo abundance from simulations in this regime. 

It would be interesting to compare the framework developed in this paper with simulations in more general models,
particularly ``single-field'' non-Gaussianity with parameters $f_{NL}^{\rm equil}$ and $f_{NL}^{\rm orth}$
\cite{Senatore:2009gt}.
The smoothed cumulants of the density field don't retain information about the ``shape'' of the $N$-point functions and one may therefore expect this framework to work for more general non-Gaussian initial conditions. (The fitting functions for the $\kappa_N$'s would of course need to be recomputed.) 
Since the log-Edgeworth mass function correctly predicts the mass function over a wide range of the $\{ \fnl,\gnl,\taunl \}$ parameter space
considered in this paper, one has increased confidence that it may be more applicable to more general non-Gaussian
initial conditions.
However, we defer a detailed comparison to future work.

\subsection*{Acknowledgements}

We thank Vincent Desjacques, Doug Rudd, Sarah Shandera, Ravi Sheth, David Spergel, and Matias Zaldarriaga for helpful discussions.
M.~L.~is supported as a Friends of the Institute for Advanced Study Member and by the NSF though AST-0807444.
K.~M.~S.~is supported by a Lyman Spitzer fellowship in the Department of Astrophysical Sciences at Princeton University.
Simulations in this paper were performed at the TIGRESS high performance computer center at Princeton University which is jointly 
supported by the Princeton Institute for Computational Science and Engineering and the Princeton University Office of Information Technology.

\bibliographystyle{ieeetr}
\bibliography{mass_function}

\appendix

\section{Calculating cumulants}
\label{app:cumulantscalc}
In \S\ref{ssec:cumulants}, we quoted fitting functions for the following cumulants:
the $\bigoh(\fnl)$ contribution to $\kappa_3$, the $\bigoh(\gnl)$ contribution to $\kappa_4$, and
the $\bigoh(\taunl)$ contributions to $\kappa_2$ and $\kappa_4$.
In this appendix, we describe our procedure for obtaining these results,
via a combination of analytic and Monte Carlo methods.

\subsection{Convergent cumulants: $\kappa_3(M)_{\fnl}$ and $\kappa_4(M)_{\gnl}$}
\label{ssec:convergent_cumulants}

We first calculate $\kappa_3(M)_{\fnl}$ and $\kappa_4(M)_{\gnl}$.
These are the cumulants which do not diverge in the infinite-volume limit.
In these cases, we can compute the cumulant efficiently by numerical integration,
as we now explain.\footnote{Note that we use a notation in which contributions of 
different orders are distinguished by subscripts,
e.g.~the first line of Eq.~(\ref{eq:k4fit}) is denoted $\kappa_4(M)_{\gnl}$ and 
the second line is denoted $\kappa_4(M)_{\taunl}$.}

The smoothed linear overdensity field $\delta_M$ and the curvature $\Phi$ are related by
\be
\delta_M(\x) = \int \frac{d^3\k}{(2\pi)^3} \alpha_M(k) \Phi(\k) e^{i\k\cdot\x}
\ee
where we have introduced the notation $\alpha_M(k) = W(kR(M)) \alpha(k)$.
(We have ignored redshift dependence, since it will eventually drop out when we
compute reduced cumulants of the form
$\kappa_N(M) = \langle \delta_M^N \rangle_{\rm conn.} / \langle \delta_M^2 \rangle^{N/2}$.)

The expectation value $\langle \delta_M(\x)^3 \rangle_{\fnl}$ can be written:
\be
\langle \delta_M(\x)^3 \rangle_{\fnl} =
  6 \fnl \int \frac{d^3\k_1}{(2\pi)^3} \frac{d^3\k_2}{(2\pi)^3} \frac{d^3\k_3}{(2\pi)^3}
       \alpha_M(k_1) \alpha_M(k_2) \alpha_M(k_3) P_\Phi(k_1) P_\Phi(k_2) (2\pi)^3 \delta_D(\k_1+\k_2+\k_3)\,.
\ee
The 6D integral can be rewritten in a more tractable form as follows.
We write the delta function as an integral $(2\pi)^3 \delta_D(\sum\k_i) = \int d^3\r\, e^{-i\sum\k_i \cdot {\bf r}}$
and introduce the notation:
\ba
\tilde \alpha_M(r) &=& \int \frac{d^3\k}{(2\pi)^3} \alpha_M(k) e^{i\k\cdot\r} \nn \\
    &=& \int_0^\infty \frac{dk}{2\pi^2} k^2 \alpha_M(k) j_0(kr) \\
\beta_M(r) &=& \int \frac{d^3\k}{(2\pi)^3} \alpha_M(k) P_\Phi(k) e^{i\k\cdot\r} \nn \\
    &=& \int_0^\infty \frac{dk}{2\pi^2} k^2 \alpha_M(k) P_\Phi(k) j_0(kr)
\ea
where $j_0(kr)=\sin(kr)/(kr)$ and we have used spherical symmetry to write each Fourier transform as a 1D integral.
Then
\ba
\langle \delta_M(\x)^3 \rangle_{\fnl} 
  &=& 6 \fnl \int_0^\infty dr\, 4\pi r^2 \tilde \alpha_M(r) \beta_M(r)^2
\ea
and $\sigma(M)^2 = \int dr\, 4\pi r^2 \alpha_M(r) \beta_M(r)$, so that the reduced
cumulant is given by:
\ba
\kappa_3(M)_{\fnl} = 6 \fnl \frac{\int_0^\infty dr\, 4\pi r^2\tilde \alpha_M(r) \beta_M(r)^2}{\left( 
  \int_0^\infty 4\pi r^2 \tilde\alpha_M(r) \beta_M(r) \right)^{3/2}}
\ea
and in this form, the cumulant can be computed efficiently, as a sequence of 1D integrals. We note that this procedure can be extended to find an expression for the $\fnl^{N-2}$ contribution to $\kappa_N$ in terms of a sequence of 1D integrals. 

A similar calculation can be done for the cumulant $\kappa_4(M)_{\gnl}$:
\ba
\kappa_4(M)_{\gnl}
  &=& \frac{24 \gnl}{\sigma(M)^4} \int \frac{d^3\k_1}{(2\pi)^3} \frac{d^3\k_2}{(2\pi)^3} \frac{d^3\k_3}{(2\pi)^3} \frac{d^3\k_4}{(2\pi)^3} 
      \alpha_M(k_1) \alpha_M(k_2) \alpha_M(k_3) \alpha_M(k_4) \nn \\
  && \hspace{2cm} \times P_\Phi(k_1) P_\Phi(k_2) P_\Phi(k_3) (2\pi)^3 \delta_D(\k_1+\k_2+\k_3+\k_4) \nn \\
  &=& 24 \gnl \frac{\int_0^\infty dr\, 4\pi r^2\tilde \alpha_M(r) \beta_M(r)^3}{\left(
  \int_0^\infty 4\pi r^2\tilde \alpha_M(r) \beta_M(r) \right)^2}\, .
\ea

\subsection{Infrared-divergent cumulants: $\kappa_2(M)_{\taunl}$ and $\kappa_4(M)_{\taunl}$}
\label{ssec:divergent_cumulants}

The cumulants $\kappa_2(M)_{\taunl}$ and $\kappa_4(M)_{\taunl}$ are formally divergent in the limit
where the simulation volume $L^3$ goes to infinity, or equivalently as integrals over wavenumbers $\k_i$
are computed with lower cutoffs $k_{\rm min} \rightarrow 0$.
This divergence can be understood physically as follows (for more discussion see \cite{Boubekeur:2005fj}).
As $k_{\rm min} \rightarrow 0$, new long-wavelength modes of $\Phi_G$ ``appear'', which can couple to modes at a fixed
physical scale $k$ (via the $\Phi_G^2$ term in Eq.~(\ref{eq:fnl_def})) and generate divergent contributions to $\Phi(\k)$.
For example, the power spectrum $P_\Phi(k)$ contains a log-divergent term proportional to $\fnl^2$.
This divergence can be regulated by fixing a box with finite volume $L^3$ and requiring that $\langle\Phi_G\rangle = 0$,
where the mean $\langle \cdot \rangle$ is defined by integrating over the box.
For a fixed $L$, all cumulants will be finite, but some contributions diverge as $\ln(L)$ as $L\rightarrow\infty$.

Our perspective is that $\fnl$ models are only defined if a box size $L$ is also specified.
When comparing with $N$-body simulations, we will choose $L$ to be the size of the simulation box.
This choice is consistent with the way the $N$-body initial conditions are generated: we do not simulate power in the DC mode of the
simulation box, and this corresponds to regulating the divergence using a volume $L^3$.
When comparing to observations, we suggest choosing $L$ to be large enough to enclose the Hubble volume.

The correction to the variance $\kappa_2(M)_{\taunl}$ is given by
\be
\kappa_2(M)_{\taunl} 
  = 2 \frac{\taunl}{(6/5)^2} \frac{1}{\sigma(M)^2} \int \frac{d^3\k_1}{(2\pi)^3} \frac{d^3\k_2}{(2\pi)^3} 
     \alpha_M(|\k_1+\k_2|)^2 P_\Phi(k_1) P_\Phi(k_2) \label{eq:kappa2_ir}
\ee
and this integral diverges in the limits $k_1\rightarrow 0$ and $k_2\rightarrow 0$.  In a finite box of volume $L^3$, it can be
regulated by replacing $\int \frac{d^3\k}{(2\pi)^3} \rightarrow \frac{1}{L^3} \sum_{\k\ne 0}$, where the sum ranges over discrete
Fourier modes (i.e.~modes of the form $(k_x,k_y,k_z) = (2\pi n_x/L, 2\pi n_y/L, 2\pi n_z/L)$ where the $n_i$ are integers).

Next we consider the $\tau_{NL}$ contribution to the kurtosis,
\ba
\kappa_4(M)_{\taunl}
  &=& 48 \frac{\taunl}{(6/5)^2} \frac{1}{\sigma(M)^4} \int \frac{d^3\k_1}{(2\pi)^3} \frac{d^3\k_2}{(2\pi)^3} \frac{d^3\k_3}{(2\pi)^3} \frac{d^3\k_4}{(2\pi)^3} 
      \alpha_M(k_1) \alpha_M(k_2) \alpha_M(k_3) \alpha_M(k_4) \nn \\
   && \hspace{2.5cm} \times P_\Phi(|\k_1+\k_2|) P_\Phi(k_1) P_\Phi(k_3) (2\pi)^3 \delta_D(\k_1+\k_2+\k_3+\k_4)\,,
\ea
this integral contains a divergence as the internal momentum ${\bf q}=(\k_1+\k_2)$ approaches zero.
To see this explicitly, we change variables:
\ba
\langle \delta_M(\x)^4 \rangle_{\taunl} 
   = 48 \frac{\taunl}{(6/5)^2} \frac{1}{\sigma(M)^4} \int \frac{d^3\q}{(2\pi)^3} P_\Phi(q)
        \left[ \int \frac{d^3\k}{(2\pi)^3} \alpha_M(k) \alpha_M(|\q-\k|) P_\Phi(k) \right]^2  \label{eq:kappa4_ir}
\ea
in the $q\rightarrow 0$ limit, the expression in brackets approaches $\sigma(M)^2 = \int \frac{d^3\k}{(2\pi)^3} \alpha_M(k)^2 P_\Phi(k)$,
a nonzero (and finite!) value.
The divergence can be regulated by replacing $\int \frac{d^3\q}{(2\pi)^3} \rightarrow L^3 \sum_{\q\ne 0}$.

It is useful to calculate the infrared divergent part of these cumulants, i.e.~the leading behavior in the infinite-volume limit.
In both cases, this will follow from analyzing the infared divergent part of the integral $\int \frac{d^3\k}{(2\pi)^3} P_\Phi(k)$,
after regulating by replacing the integral by a discrete sum over Fourier modes in a finite volume $L^3$.
For a scale-invariant power spectrum of the form $P_\Phi(k) = 2\pi^2 \Delta_\Phi^2/k^3$, it is easy to see that
\be
\int \frac{d^3\k}{(2\pi)^3} P_\Phi(k) = \Delta_\Phi^2 \ln(L) + \mbox{(finite)}  \label{eq:ir1}\,.
\ee
Comparing with Eqs.~(\ref{eq:kappa2_ir}) and~(\ref{eq:kappa4_ir}), we get the IR-divergent terms:
\ba
\kappa_2(M)_{\taunl} &=& 4 \frac{\taunl}{(6/5)^2} \Delta_\Phi^2 \ln(L) + \mbox{(finite)} \\
\kappa_4(M)_{\taunl} &=& 48 \frac{\taunl}{(6/5)^2} \Delta_\Phi^2 \ln(L) + \mbox{(finite)}  \label{eq:ir_divergent_terms}\,.
\ea
Note that the IR-divergent part is independent of $M$ in both cases.

The preceding expressions assume scale invariance for simplicity but if $n_s < 1$, we find empirically that an excellent
fitting function for the IR-divergent behavior is given by:
\be
\int \frac{d^3\k}{(2\pi)^3} P_\Phi(k) = \left( \frac{k^3 P(k)}{2\pi^2} \right)_{k=4.67/L} \frac{(L/L_0)^{n_s-1}-1}{n_s-1} + \mbox{(finite)}
\ee
which agrees with Eq.~(\ref{eq:ir1}) in the limit $n_s\rightarrow 1$.

\subsection{Monte Carlo simulations}
\label{ssec:monte_carlo}

We now describe a method for efficiently estimating cumulants for a fixed box size,
via Monte Carlo simulations of the density field.

In each Monte Carlo realization, we simulate a Gaussian initial curvature $\Phi_G$ and define
fields $\delta_M$, $\delta^*_M$, $\delta^{**}_M$ by
\ba
\delta_M(\k) &=& \alpha_M(k) \int d^3\x\, e^{-i\k\cdot \x} \Phi_G(\x) \nn \\
\delta^*_M(\k) &=& \alpha_M(k) \int d^3\x\, e^{-i\k\cdot \x} \Phi_G(\x)^2 \nn \\
\delta^{**}_M(\k) &=& \alpha_M(k) \int d^3\x\, e^{-i\k\cdot \x} (\Phi_G(\x)^3 - 3 \langle \Phi_G^2 \rangle \Phi_G(\x))\,.
\ea
The field $\delta_M$ represents the linear density field smoothed on mass scale $M$,
and the fields $\delta^*_M$, $\delta^{**}_M$ represent non-Gaussian contributions of $\fnl$-type
or $\gnl$-type respectively.

We then estimate cumulants by
\ba
\kappa_3(M)_{\fnl} &=& 3 \fnl \frac{\langle \delta_M(\x)^2 \delta^*_M(\x) \rangle}{\langle \delta_M(\x)^2 \rangle^{3/2}} \nn \\
\kappa_4(M)_{\gnl} &=& 4 \gnl \frac{\langle \delta_M(\x)^3 \delta^{**}_M(\x) \rangle}{\langle \delta_M(\x)^2 \rangle^2} \nn \\
\kappa_2(M)_{\taunl} &=& \frac{\taunl}{(6/5)^2} \frac{\langle \delta^*_M(\x)^2 \rangle}{\langle \delta_M(\x)^2 \rangle} \nn \\
\kappa_4(M)_{\taunl} &=& 6 \frac{\taunl}{(6/5)^2} \frac{\langle \delta_M(\x)^2 \delta^*_M(\x)^2 \rangle}{\langle \delta_M(\x)^2 \rangle^2}
\ea
where $\langle \cdot \rangle$ denotes an average over MC realizations and also over position $\x$ within each realization.

This Monte Carlo scheme isolates contributions of a given order (e.g.~the order-$\fnl$ contribution to $\kappa_3(M)$ is estimated
without any contribution from the order-$\fnl^3$ term) and results in good computational efficiency.
For example, one can get reasonable-looking results after a single Monte Carlo simulation.

\subsection{Fitting functions}

The convergent cumulants $\kappa_3(M)_{\fnl}$ and $\kappa_4(M)_{\gnl}$ can be calculated either
by numerical integration (\S\ref{ssec:convergent_cumulants}) or by Monte Carlo simulations of the
density field (\S\ref{ssec:monte_carlo}).  We checked that the two agree, and find that the fitting
functions given previously in Eq.~(\ref{eq:k3fit}) and the first line of Eq.~(\ref{eq:k4fit}) are excellent approximations 
(see Fig.~\ref{fig:cumulants}).

The divergent cumulants $\kappa_2(M)_{\taunl}$ and $\kappa_4(M)_{\taunl}$ are more subtle since
dependence on the box size $L$ must be included in the fitting functions.
In these cases, the IR-divergent pieces can be calculated in closed form (\S\ref{ssec:divergent_cumulants}),
but calculating the precise values of the cumulants at a finite value of $L$ is awkward since the integrals
must be regulated by summing over a discrete set of Fourier modes.
We obtain fitting functions using a hybrid approach: we use the Monte Carlo scheme from \S\ref{ssec:monte_carlo}
with fixed box size $L_0 = 1600$ $h^{-1}$ Mpc, and find an empirical fitting function for the $M$ dependence.
We then add a term proportional to $\ln(L/L_0)$ with prefactor chosen to match the IR-divergent term calculated
in Eq.~(\ref{eq:ir_divergent_terms}).
The fitting functions for $\kappa_2(M)_{\taunl}$ and $\kappa_4(M)_{\taunl}$ given previously in
Eq.~(\ref{eq:k2fit}) and the second line of Eq.~(\ref{eq:k4fit}) were obtained by this procedure.
As a check, we find that these fitting functions are excellent approximations for box sizes $L_0/4 \le L \le L_0$.
Since they have been constructed in a way which guarantees correct asymptotic behavior as $L\rightarrow \infty$,
we anticipate that they can be safely extrapolated to $L \gg L_0$, where doing simulations would be
computationally prohibitive.
This property is important since $L$ should be chosen to be Hubble-sized when comparing
with observations, as discussed in \S\ref{ssec:cumulants}.

\end{document}